\documentclass[reprint,prx, longbibliography]{revtex4-1}

\usepackage{siunitx, amsmath, amssymb}

\DeclareSIUnit[number-unit-product = \;]
	\oersted{Oe}
    
\usepackage[version=3]{mhchem}
\usepackage[inline]{enumitem}
\usepackage{graphicx}
\usepackage{dcolumn}
\usepackage{bm}
\usepackage{color}

\graphicspath{ {figures/}}

\begin{document}

\title{Broadband Optical Detection using the Spin Seebeck Effect} 

\author{Subash~Kattel$^1$}
\author{Joseph~R.~Murphy$^1$}
\author{David~Ellsworth$^2$}
\author{Jinjun~Ding$^2$}
\author{Tao~Liu$^2$}
\author{Peng~Li$^2$}
\author{Mingzhong~Wu$^2$}
\author{William~D.~Rice$^1$}
 \email{wrice2@uwyo.edu.}
\affiliation{$^1$Department of Physics and Astronomy, University of Wyoming, Laramie, Wyoming 82071, USA}
\affiliation{$^2$Department of Physics, Colorado State University, Fort Collins, Colorado 80523, USA}

\date{\today}

\begin{abstract}
The generation, control, and detection of spin currents in solid-state devices are critical for Joule-heating minimization, spin-based computation, and electrical energy generation from thermal gradients.  Although incorporation of spin functionality into technologically important architectures is still in its infancy, advantages over all-electric devices are increasingly becoming clear.  Here, we utilize the spin Seebeck effect (SSE) in Pt/Y$_3$Fe$_5$O$_{12}$ devices to detect light from 390 to 2200~nm. We find the device responsivity is remarkably flat across this technologically important wavelength range, closely following the Pt absorption coefficient.  As expected from a SSE-generation mechanism, we observe that the photovoltage and Pt heating dynamics are in strong agreement.  To precisely determine the optically created thermal gradient produced from a point-like heat source, we introduce a field-modulation method for measuring the SSE.  Our results show broadband optical detection can be performed with devices based solely on spin current generation and detection. 
\end{abstract}



\maketitle


\section{\label{sec:introduction}Introduction}

Spin current generation, detection, transport, and manipulation are key components of a new generation of spin-based devices that have both spin and electrical characteristics~\cite{JungwirthNatureMater2012, BauerNatureMater2012}.  Unlike traditional, all-electrical devices, these architectures utilize a flow of spins (i.e., a spin current density, $\mathbf{J}_s$) to transmit information and/or energy in lieu of the carrier charge~\cite{WolfScience2001, ZuticRMP2004}.  Given the weak interactions between most types of matter and carrier spins, pure spin currents transmit energy significantly more efficiently than charge currents, which unavoidably incur energetic losses in non-superconducting materials.  However, the magnetic nature of spin also means that its incorporation into traditional charge-based devices remains a technologically significant hurdle.  Moreover, it continues to be unclear that utilizing the carrier spin (instead of their charge) actually produces a marked performance enhancement in device performance except in selected cases, such as giant magnetoresistance, or in magneto-optical devices~\cite{NafradiNatureComm2016}.

To explore the advantages that pure spin currents (i.e., a net spin flow without net charge movement) may have over traditional optoelectronics, researchers have focused on three main ways to create $\mathbf{J}_s$: the spin Seebeck effect (SSE)~\cite{UchidaNature2008}, spin pumping~\cite{AzevedoJAP2005, SaitohAPL2006}, and the spin Hall effect~\cite{HirschPRL1999, KatoScience2004, ValenzuelaNature2006, JungwirthNatureMater2012}.  In particular, the SSE has gained substantial attention of the last few years as a way to convert thermal energy to electrical energy (i.e., spin caloritronics)~\cite{KiriharaNatureMater2012}.  Like the electrical Seebeck effect, in which a thermal gradient, $\nabla T$, produces a current density in materials with itinerant charge carriers~\cite{UchidaJAP2012}, the SSE occurs when a thermal gradient produces a pure spin current.  However, unlike its electrical analog, the SSE has been generated not just in ferromagnetic metals~\cite{UchidaNature2008} and semiconductors~\cite{JaworskiNatureMater2010}, but also in magnetic insulators~\cite{UchidaAPL2010, UchidaAPL2010_2, UchidaNatureMater2010}.   

Beyond these three main pure-spin-current generation mechanisms, recent work by Ellsworth et al.~\cite{EllsworthNaturePhys2016} has suggested a fourth pure-spin-current-generation pathway called the photo-spin-voltaic (PSV) effect~\cite{ZuticMaterTrans2003}, which produces $\mathbf{J}_s$ from the unequal dephasing of spin-polarized photogenerated electrons and holes in Pt. Unlike the other pure-spin-current-generation mechanisms, the PSV effect only occurs in the non-magnetic metal top layer and has only been observed via optical excitation.  Given that both the longitudinal SSE (LSSE) and PSV effect are ultimately measured via a voltage across the non-magnetic metal top layer, the all-optical nature of the PSV effect makes its empirical signatures difficult to distinguish from the optically generated LSSE.  Although a spin current density can be generated in a wide variety of magnetic materials through the aforementioned mechanisms, measuring it primarily relies on either the inverse spin Hall effect (ISHE)~\cite{AzevedoJAP2005, SaitohAPL2006, KimuraPRL2007, BartellPRAppl2017}, which produces an electrical voltage from a spin current via spin-orbit coupling, or polarization-sensitive optical detection, which utilizes out-of-plane magnetization to alter the incoming light polarization~\cite{ChoiNatureComm2014, KimlingPRL2017, McLaughlinPRB2017}.

In this work, we utilize the bulk SSE in Pt/Y$_3$Fe$_5$O$_{12}$ (YIG) bilayer devices to detect light from 390 to 2200 nm, demonstrating that pure-spin-current generation can be used for broadband optical detection.  We show that the spin current-generated ISHE voltage, $V_{\rm ISHE}$, is nearly featureless over this ultrabroadband optical range, with a responsivity, $\Re{}(\lambda)$, which follows the absorption coefficient of Pt.  In contrast to previous work~\cite{EllsworthNaturePhys2016}, we find the dynamical response of the device matches closely with the Pt thermal behavior, which suggests that the SSE is the underlying detection mechanism.  Finally, we introduce an amplitude-modulated technique for SSE detection, which we use to measure $\nabla T$ due to the optical heating from the Pt film.  From our determination of $\nabla T$, along with the measured ISHE voltage gradient, we estimate the LSSE coefficient, $S_{\rm LSSE}$ ($=\frac{V_{\rm ISHE}}{\nabla T}$), for this Pt/YIG configuration to be $60\pm$7.8~nV/K.  These results suggest that featureless, broadband photodetection can be readily achieved by thermally generated spin currents, thus avoiding spectrally limited photocarrier creation in semiconductors and showing strong similarities to thermoelectric photodetectors.

\begin{figure}
\includegraphics[width=0.5\textwidth]{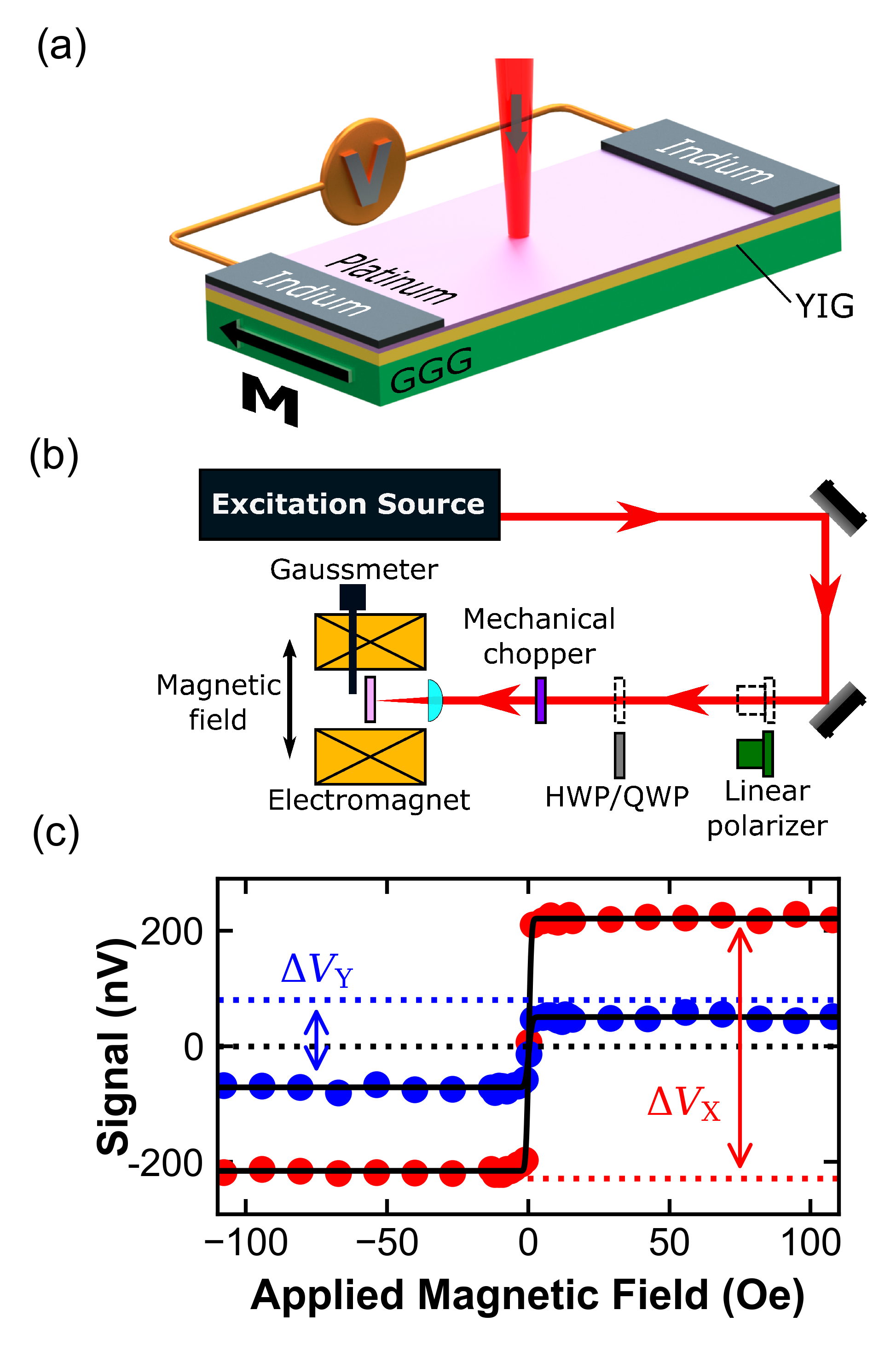}
	\caption{Pt/YIG device layout and photovoltage measurement. 
	(a) The non-magnetic metal (Pt)/magnetic insulator (YIG) bilayer structure on GGG.
	For a given device, the thickness of the Pt metal layer ranges from 2 to 6~nm.
    A magnetic field is applied perpendicular to the incident light direction producing an in-plane YIG magnetization, $\mathbf{M}$. 
    (b) A schematic of the experimental setup used to collect the steady-state excitation signal.  Depending on the experiment, a linear polarizer and either a half-wave plate (HWP) or quarter-wave plate (QWP) were added to the optical excitation path.
    (c) Field-dependent in- and out-of-phase signals, $V_{\rm X}$ and $V_{\rm Y}$, respectively, measured using a lock-in amplifier. 
    The solid lines are error function fits to the data. 
    \label{Fig:experimental_setup}}
\end{figure}

\section{\label{sec:Results and Discussion}Results and Discussion}

In this work, we use bilayer devices prepared using the procedure described in Ellsworth et al.~\cite{EllsworthNaturePhys2016}. Briefly, a layer of YIG is grown via liquid phase epitaxy on a Gd$_3$Ga$_5$O$_{12}$ (GGG) substrate with a thickness of either 80~nm or 15~$\mu$m depending on the sample series. Next, RF magnetron sputtering overcoats the YIG with Pt.  We use Pt/YIG devices that had a Pt film thickness, $t_{\rm Pt}$, which ranges from 2 to 6~nm; a diagram of this heterostructure is shown in Fig.~\ref{Fig:experimental_setup}(a). In the first Pt/YIG/GGG device series (YIG thickness = 15~$\mu$m), the 500~$\mu$m-thick GGG is polished only on the side on which the YIG layer is deposited; in the second device set (YIG thickness = 80~nm), the GGG substrate is polished on both sides, which allows optical excitation from either face of the device.

Figure~\ref{Fig:experimental_setup}(b) schematically shows the experimental setup used to obtain the majority of the results in this work. The measurements were conducted in the LSSE geometry~\cite{UchidaAPL2010, BartellPRAppl2017}:  a uniform magnetic field is applied in the plane of the device perpendicular to the optically induced $\nabla T$, creating a $\mathbf{J}_s$ that is proportional to the local YIG magnetization, $\mathbf{M}$.  As will be shown later, we use the ISHE to detect $\mathbf{J}_s$, which creates a spin-dependent transverse electric field that is orthogonal to both $\nabla T$ and $\mathbf{M}$~\cite{BartellPRAppl2017}.  The exact origin of the LSSE, either from a thermal gradient across the YIG~\cite{UchidaAPL2010} (bulk SSE) or a difference of interfacial temperatures between the Pt and YIG~\cite{KimlingPRL2017} (interfacial SSE), cannot be unambiguously determined from either experimental technique that we use in this manuscript; however, as shown later, we are very likely observing the bulk LSSE.  Although the LSSE geometry is straightforward to implement, its configuration and our use of thin Pt ($<$10~nm) allows other mechanisms, such as the anomalous Nernst effect~\cite{SakaiNaturePhys2016} and magnetic proximity effects~\cite{HuangPRL2012}, to occur.  However, previous work done using directional magnetization~\cite{KikkawaPRB2013} and Pt/Cu/YIG and Au/Cu/YIG structures~\cite{MiaoAIPAdvances2016} strongly suggests that the LSSE dominates over these effects in Pt/YIG bilayers.

We measure the optically generated $V_{\rm ISHE}$ between two pressed indium contacts using a lock-in amplifier (LIA) triggered by an optical chopper.  
The resulting signal from a standard field sweep ($-$100 to $+$100~Oe), and the corresponding error function fit, is shown in Fig.~\ref{Fig:experimental_setup}(c).  We clearly observe that as the field is swept, the device voltage rapidly rises from negative to positive values between $-$30 and $+$30~Oe, reflecting the change in the YIG magnetization.  We call the in-phase difference in signal level between opposite field directions $2\Delta V_{\rm X}$, which is defined as $\left[V_{{\rm X},{\rm H}_+}-V_{{\rm X},{\rm H}_-}\right] \approx 2V_{{\rm X},{\rm H}_+} \approx 2V_{{\rm X},{\rm H}_-}$~\cite{AgrawalPRB2014, BartellPRAppl2017}; the asymmetry in the field-dependent signal that is sometimes observed ($V_{{\rm X},{\rm H}_+} \neq V_{{\rm X},{\rm H}_-}$) is related to an unintentional tilt of the device in the field.  $\frac{1}{2}\Delta V_{\rm X}$ is reported as the device signal throughout this work, except for the last experiment.  

\subsection{Detection via the inverse spin Hall effect}

In distinct contrast to photoconductivity, where photogenerated carriers produce a measurable voltage (or current), and optical thermoelectricity, in which a photo-induced thermal gradient produces a voltage via the (electrical) Seebeck effect, we \textit{only} observe a photovoltage in the presence of a (small) magnetic field.  As demonstrated in Fig.~\ref{Fig:experimental_setup}(c), the steady-state voltage obtained at higher applied fields shows that the YIG must be fully magnetized in order to achieve full device responsivity~\cite{UchidaAPL2010}.  More precise information can be gleaned if we pair the how the signal varies as a function of field with its angular dependence.  Specifically, if we are indeed using the ISHE to detect spin currents, then rotating the magnetic field direction, and thus the spin polarization vector, $\boldsymbol{\sigma}$, will change the electric field generated between the two contacts, $\mathbf{E}_{\rm ISHE}$~\cite{SaitohAPL2006, UchidaAPL2010}: 

\begin{equation}
-\nabla V_{\rm ISHE} = \mathbf{E}_{\rm ISHE} = \frac{1}{\sigma_c} D_{\rm ISHE}(\mathbf{J}_s \times \boldsymbol{\sigma}), 
\label{ISHE_voltage_eqn}
\end{equation}

\noindent where $\sigma_c$ is the electrical conductivity ($\mathbf{E}_{\rm ISHE} = \frac{1}{\sigma_c}\mathbf{J}_c$), and $D_{\rm ISHE}$ is the ISHE coefficient. 

\begin{figure}[h]
	\includegraphics[width=0.95\columnwidth]{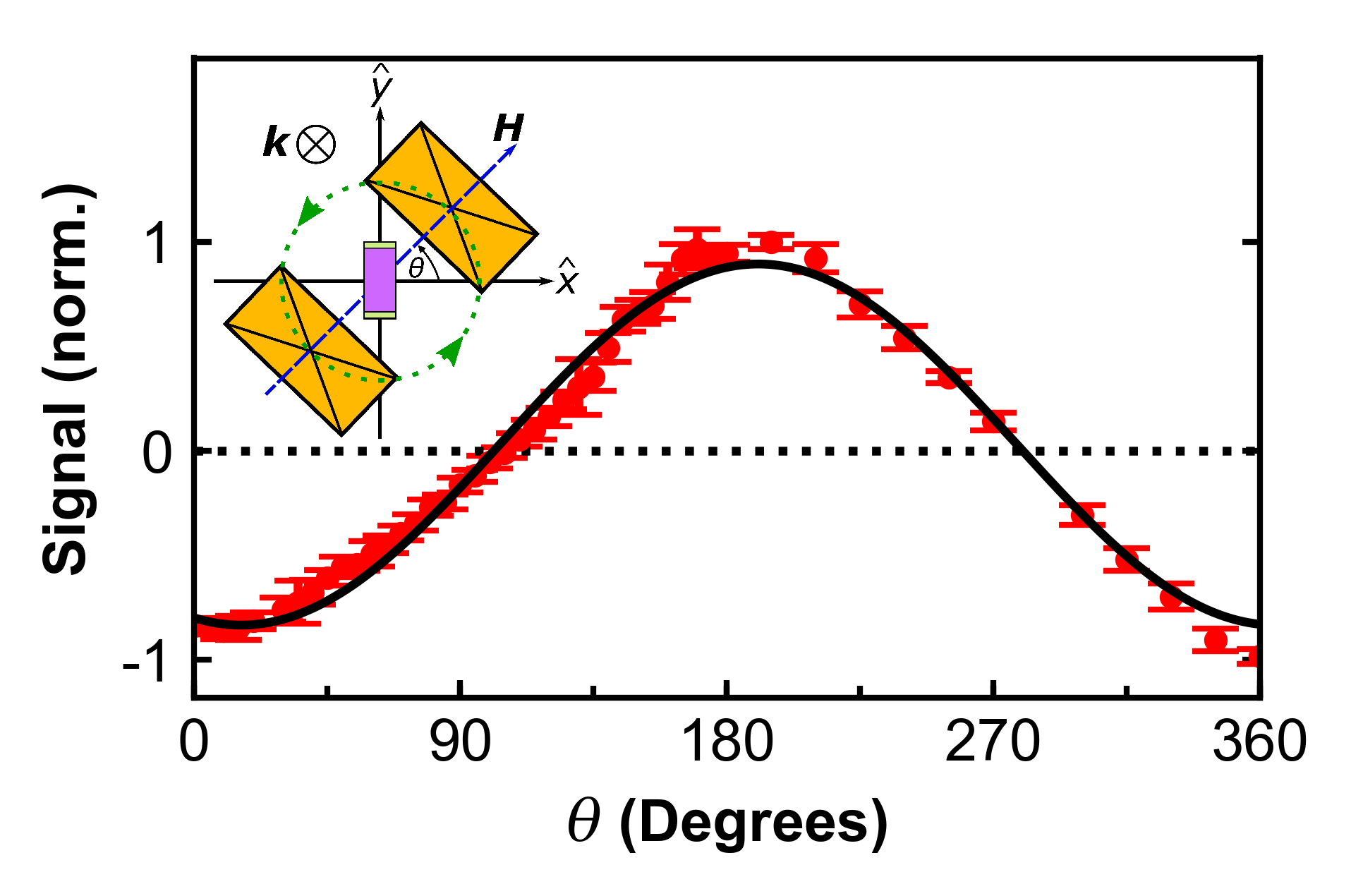}
	\caption{As the direction of magnetic field is rotated with respect to the bilayer device, the variation of the (normalized) signal exhibits a sinusoidal pattern (black fit line), which agrees well with previous ISHE observations.  Inset: Graphical depiction of the rotational field dependence experiment.  The electromagnets (orange) were rotated a full 360$^{\circ}$ about the bilayer device.  With the indium contacts aligned along the $\hat{y}$ axis, $\theta$ is defined as the angle between the applied magnetic field, $\mathbf{H}$, and the $\hat{x}$ axis. The incident light, whose direction is defined by the $\mathbf{k}$ vector, is set in the $-\hat{z}$ orientation.
	\label{Rotational_dependence_fig}}
\end{figure}

To measure this field angle dependence in Eq.~\ref{ISHE_voltage_eqn}, we placed the Pt/YIG bilayer device between the poles of an electromagnet that had the ability to rotate a full 360$^{\circ}$ around the bilayer detector (Fig.~\ref{Rotational_dependence_fig}).  For each angle of the electromagnet, $\theta$, taken with respect to the $\hat{x}$ axis, the magnetic field was scanned from $-$100~Oe to $+$100~Oe.  The field scan at each angle was then fit with an error function from which the signal was determined.  

Figure~\ref{Rotational_dependence_fig} shows the magnitude of the resulting signal from this field sweep as a function of $\theta$.  The normalized signal clearly follows a sinusoidal pattern (black fit line), which corresponds to $\boldsymbol{\sigma}$ rotating from orthogonal to $\mathbf{E}_{\rm ISHE}$ (maximum magnitude) to parallel with it (zero magnitude) and is consistent with previous works using the pure spin current detection using the ISHE~\cite{SaitohAPL2006, UchidaAPL2010, WangPRL2014,EllsworthNaturePhys2016}.  Importantly, the use of the magnetic field vector to observe a photovoltage from a Pt-based device demonstrates spin current creating functionality and distinguishes these measuremetns from a photoconductive effect in either Pt or YIG.  


\subsection{\label{sec:Spectral response of the Pt/YIG devices}Spectral response of the Pt/YIG devices}

\begin{figure*} [htbp]
	\includegraphics[width=0.95\textwidth]{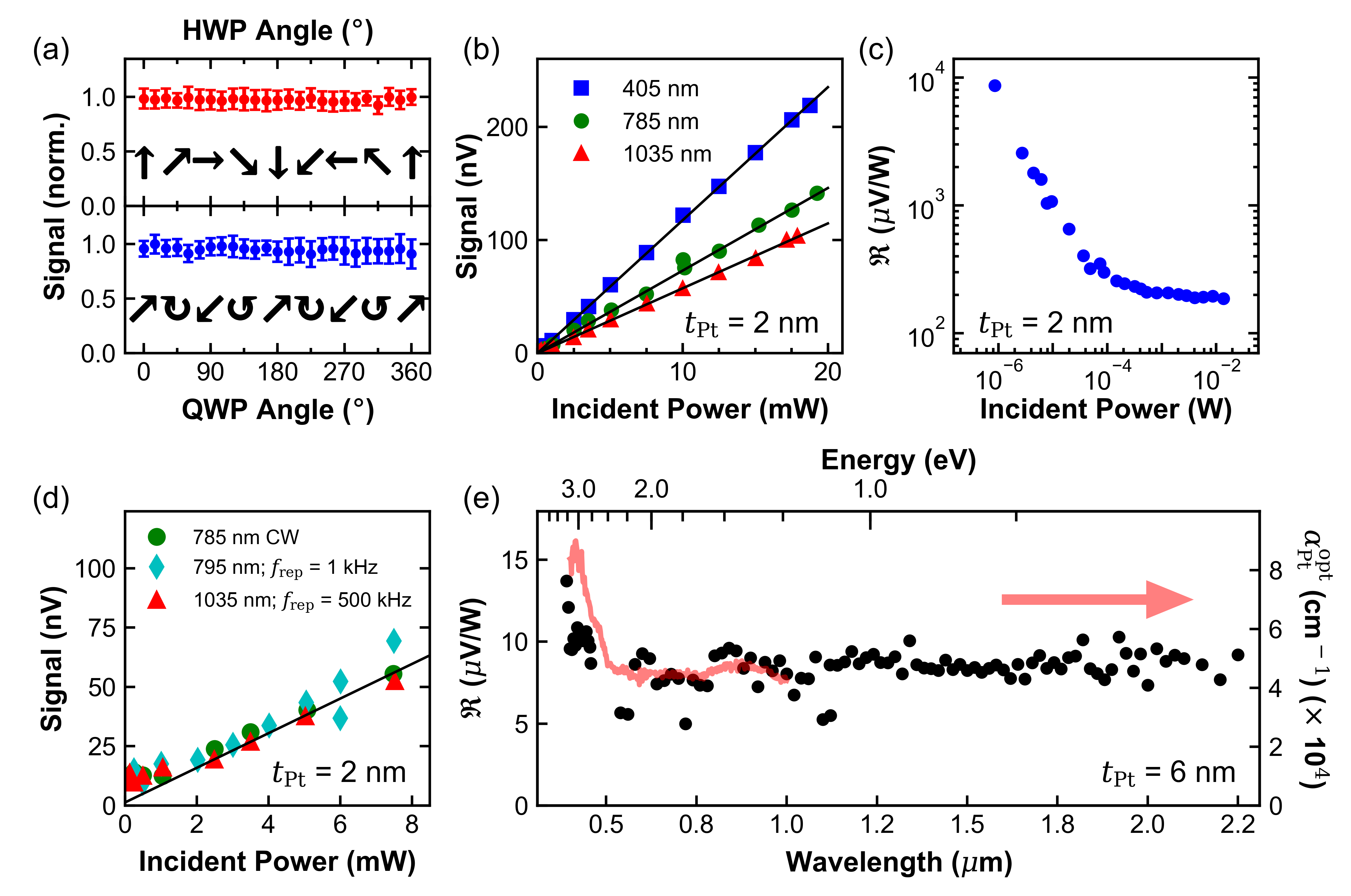}
	\caption{Optical response of the Pt/YIG device. 
	(a) In the upper panel, the normalized $V_{\rm ISHE}$ produced by the incident light shows no change as the linear polarization (half-wave plate; HWP) angle is rotated.
	Similarly, in the lower panel, we observe no change when the light polarization is changed via rotation of a quarter-wave plate (QWP).
	(b) The signal power dependence for three wavelengths shows a linear scaling, which allows us to normalize the signal by incident power (i.e., device responsivity).
	The magnitude of the signal from a 405 nm excitation is clearly larger than the signals from 785~nm and 1035~nm excitation wavelengths.
	(c) Device responsivity, $\Re{}(\lambda)$, as a function of incident power for 405 nm on a 2 nm-thick Pt/YIG device.
	(d) Voltage signal created by illumination from different pulsed (795~nm, 1 kHz repetition rate, pulse duration $\sim$180~fs; 1035~nm, 500~kHz repetition rate, pulse duration $\sim$220~fs) and CW sources as a function of average power. Regardless of repetition rate, the device signal linearly scales with the average incident power.
    	(e) The device responsivity, $\Re{}(\lambda)$, from 390 to 2200~nm (black dots) is flat from 450 to 2200~nm.
	Below 450~nm, we observe a small increase in $\Re{}$. 
	This spectral trend closely follows the measured absorption coefficient of the Pt film (red).
	\label{Fig:figure2}}
\end{figure*}

The initial findings of Ellsworth and co-workers~\cite{EllsworthNaturePhys2016} suggested that the spectral dependence of these bilayer systems may help to determine the primary spin-current-generation mechanism:  that is, distinguish between the SSE or PSV effect.  To measure device spectral response, we use a wide variety of different illumination sources, which necessitated examining how the Pt/YIG devices responded to different light polarizations, incident power, and pulse repetition rate.  The upper panel of Fig.~\ref{Fig:figure2}(a) shows that the voltage signal is constant when the linear polarization is rotated via a half-wave plate (HWP) over 360$^{\circ}$.  Similarly, despite the strong spin-orbit coupling of Pt~\cite{KimuraPRL2007}, the device signal remains unchanged when the ellipticity of the light polarization is tuned from linear to circular with a quarter-wave plate (QWP).  The lack of polarization dependence is significant not just from a technological point of view, but it also enables our implementation of the double-sided illumination experiment discussed later.

To compare the device response obtained at different wavelengths, we needed to verify that we could normalize the signal by incident power. In Fig.~\ref{Fig:figure2}(b), we find that the signal scales linearly for several different wavelengths.  This linearity allows us to define a device responsivity, $\Re{}$, for a given wavelength, $\lambda$:  $\Re{}(\lambda) = V_{\rm ISHE}(\lambda)/P_{\rm incident}(\lambda)$, where $P_{\rm incident}(\lambda)$ is the incident power at $\lambda$.  Figure ~\ref{Fig:figure2}(c) shows the measured $\Re{}$ as a function of incident power across over four orders of magnitude for the 2 nm-thick Pt device.  Despite the large change in incident power, the responsivity drop is less than a factor of 100, which compares favorably with power-dependent $\Re{}$ trends in similarly novel device structures, such as black phosphorous carbide~\cite{TanAdvMater2018} and graphene heterostructures~\cite{LongNanoLett2016}.  

Figure~\ref{Fig:figure2}(d) shows the power-dependent voltage response from excitation sources with different pulse repetition rates.  We find that as the repetition rate of the laser is varied from continuous wave (CW) to 1 kHz to 500 kHz (pulse duration $\approx$200~fs for both pulsed excitation sources), the device response is exactly the same.  This behavior suggests the mechanism for generating a photovoltage is based on thermal effects, since the signal only depends on the average incident power.  We note that previous researchers have shown spin accumulation times for Pt/YIG structures on the order of 10$^{-12}$ to 10$^{-11}$ seconds~\cite{KimlingPRL2017} and magnon dynamics on the order of 10$^{-7}$ seconds~\cite{AgrawalPRB2014}, which occur on much faster timescales than the data shown here. Importantly, the similarity our observed device response between CW and ultrafast excitation (including at different repetition rates) as a function of average power strongly suggests that the interfacial SSE, which has been shown to occur between the Pt and YIG on timescales of tens of picoseconds~\cite{KimlingPRL2017}, is not influencing our steady-state observations; based on this (indirect) evidence, we believe the bulk SSE is the dominant mechanism here.

The spectral dependence of the device responsivity from 390 to 2200~nm was obtained using several different excitation sources.  In the UV region, 390 to 460~nm, a frequency-doubled, Ti:sapphire oscillator was used, while in the visible and near-infrared regimes (540 to 2200~nm), we utilized a combination of second-harmonic generation, sum-frequency generation, and signal/idler beams from a Ti:sapphire-pumped, optical parametric amplifier (OPA).  Finally, we confirmed the near-infrared results using a tunable-wavelength, Ti:sapphire oscillator from 760 to 990 nm. 

To ensure that this broadband spectral dependence is accurate, we used significant filtering to eliminate spurious spectral components from the illumination sources.  
This operation was especially critical for the OPA output, which contains multiple output wavelengths due to the nature of the optical generation mechanisms.  
Given the polarization dependence of the various frequency-mixing operations (sum-frequency, difference-frequency, and second-harmonic generation), we directed the OPA output through two linear polarizers to remove significant fractions of non-relevant wavelengths, which was followed by set of dielectric and elemental filters (germanium, e.g., for near-infrared light) to further isolate the desired $\lambda$.
To ensure the spectral purity of the incident light, we passed the light through a Princeton Instruments SP2150 monochromator and performed a lock-in amplifier-based scan using a single-channel InGaAs detector. 
These spectral scans were done every other excitation wavelength above 1100~nm.

Figure~\ref{Fig:figure2}(e) shows the spectral dependence of $\Re{}(\lambda)$ for the 6~nm-thick Pt/YIG device spanning from 390 to 2200~nm, the limits of our optical excitation range. In contrast to PSV predictions~\cite{EllsworthNaturePhys2016} and semiconductor-based photodetectors, the measured spectral responsivity is nearly \textit{featureless}:  we observe no significant changes in the value of $\Re{}(\lambda)$ or sign flips.  The red curve in Fig.~\ref{Fig:figure2}(e) displays the measured optical absorption coefficient, $\alpha^{\rm opt}_{\rm Pt}$, from 390 to 1000~nm of the 6~nm-thick Pt layer.  The matching spectral dependences of both $\Re{}(\lambda)$ and $\alpha^{\rm opt}_{\rm Pt}$ strongly suggest that the absorption of the Pt is central to the observed behavior.  

The unexpectedly broadband, flat spectral responsivity of our devices represents a remarkable improvement, both in spectral uniformity and range, over traditional semiconductor-based photodectors, as well as more exotic configurations claiming ultrabroadband responsivities, like few-layer black phosphorous~\cite{BuscemaNanoLett2014}, ferroelectric-gated MoS$_2$~\cite{WangAdvMater2015}, MoS$_2$-graphene-WSe$_2$ heterostructures~\cite{LongNanoLett2016}, graphene heterostructures~\cite{LiuNatureNano2014}, and nanotube-graphene hybrids~\cite{LiuNatureComm2015}.  The extension of the device spectral range deep into the near infrared (and possibly much farther) opens up the possibility for using this type of architecture for converting thermal radiation into electrical energy, similar to demonstrations of spin-based thermoelectrics~\cite{KiriharaNatureMater2012}. Unlike photodetectors based on photovoltage, photogating, plasmon or cavity enhancement~\cite{BandurinNatureComm2018}, or carrier multiplication/avalanching, in which either the carrier population or the carrier mobility is (optically) changed, our device has more in common with the broad spectral responsivities seen in thermally based detection mechanisms, such as photo-thermoelectrics, thermopiles, pyroelectrics, and bolometers, in which device conductance is altered via lattice temperature. However, unlike these more traditional mechanisms in which the semiconductor carrier mobility or metallic/superconductor conductivity is altered by an optically created temperature increase, our detection mechanism relies on the optical generation and electrical detection of spin currents.  This fundamentally different photodetection process is advantageous because it leverages materials with high absorptivity (e.g., heavy metals) that do not necessarily create easily-recoverable carriers, while avoiding the lossy process of carrier separation, injection, and recovery.  Although we note that the magnitude of our measured $\Re{}(\lambda)$ remains well below more-optimized, photocarrier-based architectures~\cite{LiuNatureNano2014, LiuNatureComm2015}, recent work on improving the SSE efficiency~\cite{YuasaJPhysD2018, NakataJJAP2019} show that much larger values of $\Re{}(\lambda)$ are achievable in spin caloritronic devices.

\subsection{Influence of the Platinum Layer Thickness on the Device Responsivity}

Although the Pt film is the thinnest part of the bilayer structure (thinner, in fact, than the estimated Pt skin depth of $\sim$11 nm in this wavelength range), the large absorption coefficient ($\alpha^{\rm opt}_{\rm Pt}${$\sim$}10$^{5}$~cm$^{-1}$) is the dominant per unit length absorption contribution in the system. 
As mentioned above, we used Pt/YIG bilayers with varying $t_{\rm Pt}$ to generalize our observations.  Given that total absorption increases linearly with $t_{\rm Pt}$, a naive expectation is that $V_{\rm ISHE}$ and $\Re{}$ follows this same dependence. 

\begin{figure}[hbt]
	\includegraphics[width=0.95\columnwidth]{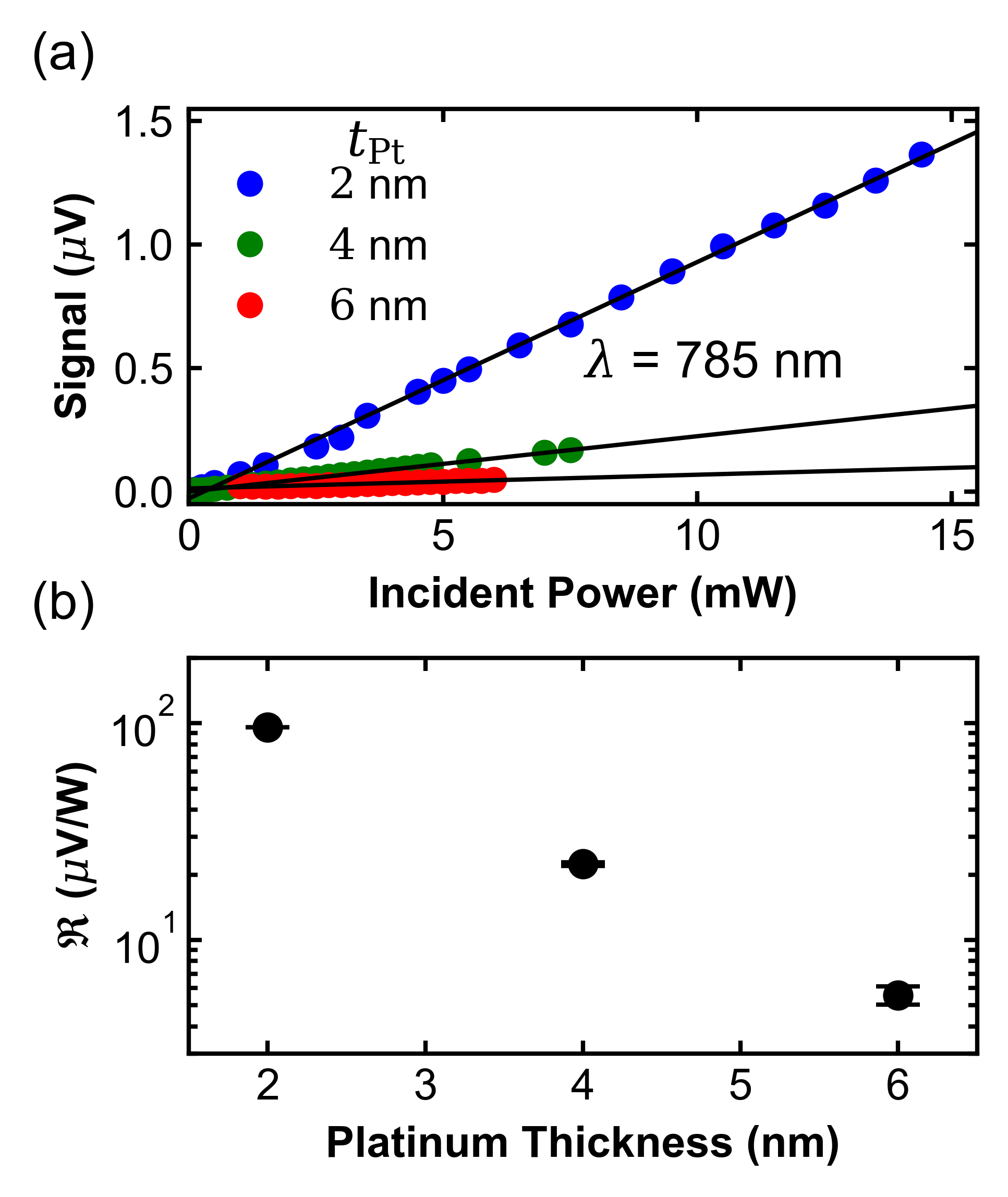}
	\caption{(a) The SSE signal at 785 nm scales linearly with incident power for different Pt film thicknesses, $t_{\rm Pt}$. (b) Responsivity, $\Re{}(\lambda =785$~nm$)$, as a function of Pt film thickness.  For the three Pt thicknesses measured, we see that as $t_{\rm Pt}$ decreases, $\Re{}$ significantly increases.
	\label{Thickness_dependence_fig}}
\end{figure}

Figure~\ref{Thickness_dependence_fig}(a) shows $V_{\rm ISHE}$ as a function of incident power for all three Pt film thicknesses.  It is immediately clear that a monotonic increase photovoltage is observed with decreasing $t_{\rm Pt}$.  A plot of $\Re{}$ as a function of $t_{\rm Pt}$ in Fig.~\ref{Thickness_dependence_fig}(b) indicates that device responsivity drops by over an order of magnitude when the Pt film thickness is increased from 2 to 6 nm.  Follow-on measurements of $\Re{}(\lambda)$ from 390 to 1600 nm show that this Pt thickness scaling holds across a broad range of wavelengths.  These results suggest that the energy absorbed per unit length, which is proportional to $\alpha^{\rm opt}_{\rm Pt}$, and the proximity to the YIG layer are two critical parameters for determining $\Re{}$.  Nevertheless, it is still surprising that as thickness is increased from 2 to 6 nm, $\Re{}$ drops by over an order of magnitude.  Previous work on changing $t_{\rm Pt}$ in Pt/YIG bilayers~\cite{WeilerPRL2013}, as well as theoretical models of how $S_{\rm SSE}$ varies with Pt thickness, show that as $t_{\rm Pt}$ decreases, $S_{\rm SSE}$ increases.  However, the surprising increase of $\Re{}$ with decreasing $t_{\rm Pt}$ may be affected via other factors, such as spin scattering at the Pt/YIG interface, differences in Pt film wetting on the YIG~\cite{AqeelJAP2014}, thermal gradients across the Pt film~\cite{SurabhiScientificRep2018}, differences in the effective absorption coefficient of the Pt~\cite{SurabhiScientificRep2018}, interfacial Rashba spin-orbit interactions, and thickness-dependent spin diffusion lengths.  Further investigation of how $\Re{}$ depends on $t_{\rm Pt}$ is ongoing.

\subsection{Conditions on the photovoltage generation}

In the previous sections, the importance of the Pt layer in the operation of the device was firmly established, while the role of the GGG, and in particular, heating effects related to its absorption of light, remained unexplored.   In this section, we isolate the role of the GGG by altering either the excitation geometry or manipulating the Pt layer.

\begin{figure*}
	\centering
	\includegraphics[width=.97\textwidth]{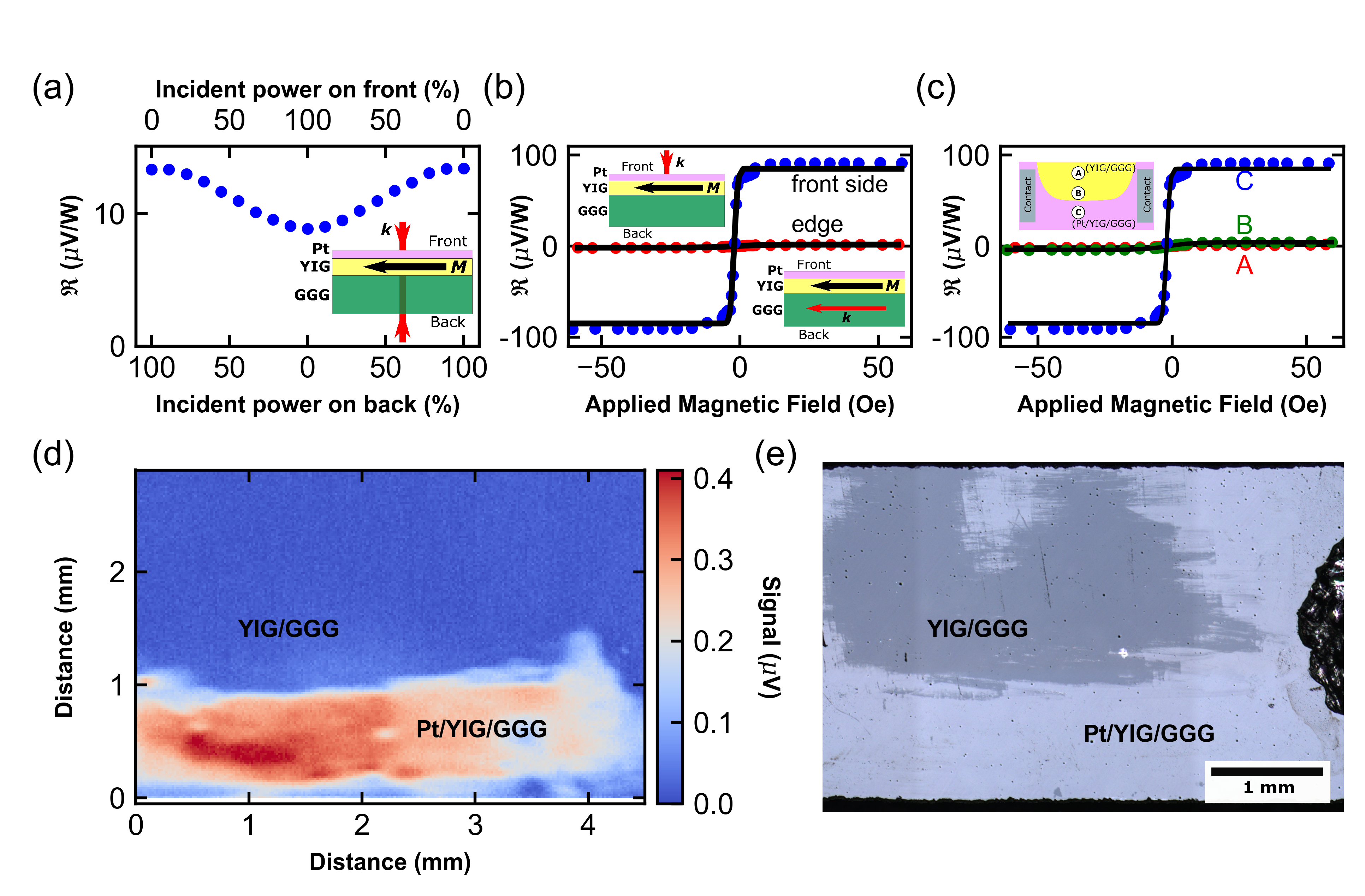}
	\caption{(a) Device responsivity as optical excitation is varied between the front and back of the device.
	Since the sign of $\Re{}(\lambda)$ does not change when the illumination direction is switched from front to back, we conclude that the optically generated thermal gradient remains the same regardless of the illumination orientation.
	(b) When a 785 nm diode laser is incident on the front side of the 2 nm-thick Pt/YIG device (blue), a large signal is observed.  In contrast, when this beam at the same power is incident on the GGG only (from the side; red), a very small voltage signal is obtained.
	(c) The signal obtained illuminating Pt/YIG/GGG (blue) and two different locations (red and green) on YIG/GGG with the Pt top layer removed.
	The solid black lines are fits to data.
    The top left inset shows the illumination spot on the sample in a region with (represented by `C') and without the Pt layer (`A' and `B').
	(d) Two-dimensional mapping of the voltage signal showing variation across a sample with partially removed Pt areas.
	(e) Optical image of the sample where the signal mapping in (c) was performed.
	\label{GGG_heating_fig}}
\end{figure*}

To see if we could produce a thermal gradient across the YIG via some manner other than illuminating the Pt top layer, we attempted to directly optically heat the GGG substrate.  In the first experiment, the GGG substrate was excited from the back side of the bilayer with the goal of flipping the sign of $\nabla T$.  While keeping the total photon number fixed, the percentage of the total power incident on the front and back was continuously varied by means of a HWP and a polarization-sensitive Glan-Taylor polarizer in front of a beamsplitter.  Although the incident light polarization states on the front and back sides of the device are different in this geometry, we showed in Fig.~\ref{Fig:figure2}(a) that $V_{\rm ISHE}$ is polarization insensitive.  
Figure~\ref{GGG_heating_fig}(a) demonstrates that as the incident power illumination is shifted, the $\Re{}$ only slightly changes.  The insignificant difference with illumination direction suggests that the GGG (and YIG) is not absorbing a substantial amount of light per unit length~\cite{Scottpssb1977, EllabbanOpticsExpr2006, BartellPRAppl2017}.  If it was, $\nabla T$ across the YIG would flip signs, thus reversing the direction of $\mathbf{J}_{s}$ and inverting the sign of $V_{\rm ISHE}$.  

As a follow-on to this experiment, we directly illuminated the GGG from the side of the device (``edge illumination'') using different excitation sources (405 and 785~nm diode lasers).  Since the front side of the device is at room temperature (i.e., no Pt heating = `cold' Pt-YIG interface), we are again interested in seeing if the temperature rise of the GGG reverses the thermal gradient across the YIG, thus flipping the sign of $V_{\rm ISHE}$.  However, this time, we are avoiding any illumination of the Pt film in an attempt to better distinguish between the PSV effect (which requires Pt optical excitation) and the SSE.  

To perform the edge-illumination experiment, we used diamond-grit abrasive paper to polish the two opposite sides of the device, which minimized optical scattering.
As the edges were lapped, we monitored the surface quality and device integrity using an optical microscope.  Light was directed through the entire length of the GGG, thus only optically heating the bottom of the YIG via the hot GGG. 
Figure~\ref{GGG_heating_fig}(b) clearly shows that the signal obtained during edge illumination is very small when compared to front-side illumination with same excitation power.  Taken together, these two direct GGG optical excitation results show that either (1) the $\nabla T$ ($\propto V_{\rm ISHE}$) across the YIG is dominated by optical heating of the Pt layer and contributions from the GGG substrate are negligible or (2) the PSV effect is actually the dominant spin current generation mechanism instead of the SSE.

In addition to optical heating of the YIG top layer via the nanometer-thick Pt film, we find that a strong photocreated signal also requires that the Pt film remain continuous and connected to the contacts.  To demonstrate this condition, we mechanically removed half of the Pt film from the 2 nm-thick Pt/YIG device using a plastic spatula.  An optical microscope was used to monitor the removal process to avoid damaging the device. 
We performed a point-by-point scan of the half-Pt device using a three-axis stage, which was automated for three-axis movement and a long working-distance microscope objective (Mitutoyu NIR 50$\times$) with a measured spot size of~20 $\mu$m.
We scanned an area of~3.3 mm $\times$ 4.5~mm with a step size of 5~$\mu$m using an excitation wavelength of 785 nm; at each location, $V_{\rm ISHE}$ was measured via a lock-in amplifier locked to a mechanical chopper modulating the light intensity.   Although 785 nm is outside the higher absorption regions of YIG and GGG, we observe similar (with single-location excitation) using a 405 nm laser.
After confirming the $V_{\rm ISHE}$ signal with a magnetic field sweep at a fixed location, the static applied field was set at 100~Oe, and the device was moved in a 2D pattern.  

Figure~\ref{GGG_heating_fig}(d) shows the result of this scan, with the regions indicated in red corresponding to the highest level of signal.
In direct comparison to this SSE signal map, an optical image of the same total area is shown in Fig.~\ref{GGG_heating_fig}(e).
The region that corresponds to the highest signal in the 2D signal map is the region where the Pt film remained continuous between the two indium contacts.
As expected, the device generates a voltage when the optical beam illuminates a region of intact Pt.  Interestingly, however, we do not see an appreciable signal when the optical excitation is on Pt regions that are disconnected from the indium contacts.  This observation, along with the measured field-angle dependence of $V_{\rm ISHE}$ (Fig.~\ref{Thickness_dependence_fig})~\cite{SaitohAPL2006, UchidaAPL2010}, confirms that we are using the ISHE to detect spin currents.  However, despite the substantial amount of data presented thus far, we still cannot conclusively determine whether the SSE and/or PSV effect is the dominant spin-current generation mechanism. \newline

\subsection{Thermal Response Dynamics}
\begin{figure}[h!]
\includegraphics[width=0.97\columnwidth]{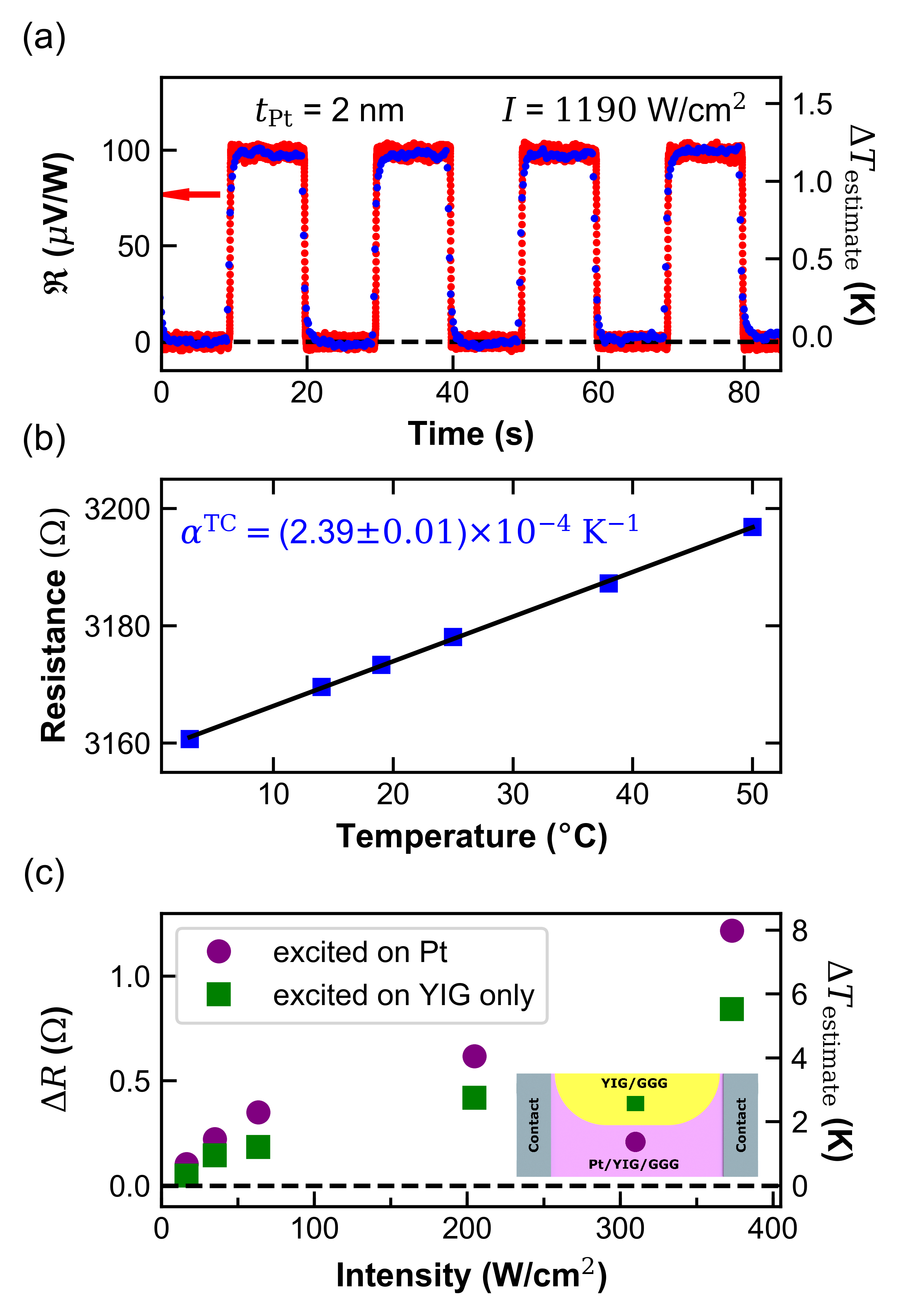}
	\caption{(a) Device responsivity (red dots) with illumination off and on in twenty-second intervals.
	The device response is faster than the measurement time resolution (10~ms).
	(Right vertical axis) Estimated temperature change (blue dots), $\Delta T_{\rm estimate}$, of the Pt film using the measured value for $\alpha^{\rm TC}_{\rm Pt}$ during the off-on illumination cycles.
	The dynamics (within our resolution) of the power-normalized signal and bulk resistivity change to optical excitation are the same.
	(b) Increase in resistance of the Pt film with the temperature. The black solid line is the linear fit to the data. The measured value of temperature coefficient $\alpha^{\rm TC}_{\rm Pt}$ was found to be 2.39$\times 10^{-4}$~K$^{-1}$ when illuminated with~1190~W/cm$^2$ using 405~nm laser as excitation source.
	(c) The Pt layer temperature rise (right y-axis), $\Delta T_{\rm estimate}$, can be estimated from the change in the Pt resistance, $\Delta R$.  As indicated by the inset, we illuminated on (purple dots) and off (green squares) the Pt/YIG bilayer.
   \label{Thermal_response_fig}}
\end{figure}

In order to help distinguish between the PSV effect and the SSE, we measured the temporal device response and compared it to the temperature dynamics of the Pt layer.  As shown earlier in Fig.~\ref{Fig:figure2}(d), the magnitude and (average) power scaling of $V_{\rm ISHE}$ did not change when the optical illumination was changed from a continuous excitation to various pulsed sources, which suggests that the signal is generated from a purely thermal heating mechanism.  A more direct measurement of this hypothesis is given in Figure~\ref{Thermal_response_fig}(a), which shows the signal response (red) of four, light-on/light-off cycles over the course of $\sim$85~seconds.  When the light is unblocked, the device immediately (to within our instrument resolution of 10~ms) generates a voltage, which is consistent with results obtained by Ellsworth and co-workers~\cite{EllsworthNaturePhys2016} using a broadband source. However, unlike these previous results, the \textit{thermal} response of the Pt to the light exactly follows the signal response.  This rapid temperature rise and equilibration of the Pt layer temperature, and its strong agreement with the device response, both suggest that we are observing the SSE rather than a photocarrier-mediated process.  

Deeper insight into the SSE mechanism in these devices can be obtained by determining the magnitude of the Pt temperature change created by the optical beam.  This estimate necessitates determining the temperature coefficient of resistance for the Pt film, $\alpha^{\rm TC}_{\rm Pt}$, which is used in the relation for temperature-dependent resistance, $R\left(T\right)$~\cite{ZhangJJAP1997}: $R\left(T\right) = R_0\left[1+\alpha^{\rm TC} \left(T-T_0\right)\right]$, where $R_0$ is the resistance at a known temperature $T_0$. In Fig.~\ref{Thermal_response_fig}(b), we show the change of $R$ as a function of temperature to obtain $\alpha^{\rm TC}_{\rm Pt} = \left(2.39\pm 0.01\right)\times 10^{-4}~{\rm K^{-1}} = 0.0736\left(5\right) \alpha^{\rm TC~bulk}_{\rm Pt}$.  The discrepancy between the bulk Pt temperature coefficient, $\alpha^{\rm TC~bulk}_{\rm Pt}$~\cite{Gale2004, ZhangJJAP1997}, and the measured temperature coefficient of our thin Pt layers, $\alpha^{\rm TC}_{\rm Pt}$, is attributable to the incomplete and fragmented nature of metallic films when the thickness is on the order of nanometers~\cite{NambaJJAP1970}.  Using this lower value for $\alpha^{\rm TC}_{\rm Pt}$, we find that the estimated Pt film temperature rise, $\Delta T_{\rm estimate}$, created by an optical intensity of 1190 W/cm$^2$ is $\sim$1.5~K.  If we assume the temperature of the bottom side of the YIG to be unchanged by the optical beam (as we empirically showed earlier), then the $\sim$1.5~K change in the Pt film is actually an estimate of $\Delta T$ across the YIG, a value that is large enough to observe a small SSE signal.  Despite the efforts we took to determine $\Delta T$ from directly measuring $\Delta R$ of the Pt layer, this estimate is not very accurate in our particular case for two reasons:  (1) the optical illumination is over only a fraction of the total Pt layer and (2) the optical beam has a Gaussian profile.  In both cases, we are deviating substantially from the situation of an evenly heated bulk slab of Pt, which is intrinsically assumed in our calculation of $\Delta T_{\rm estimate}$. 

Additionally, we also measured the change in the Pt temperature when the YIG/GGG was optical illuminated.  
In this experiment, we measured the Pt temperature as a function of optical power when the beam was on and off of the Pt.  
Although we did not measure an appreciable $V_{\rm ISHE}$ from the YIG/GGG illumination (Fig.~\ref{GGG_heating_fig}), we did observe a significant temperature change of the \textit{adjacent} Pt top layer, as shown Fig.~\ref{Thermal_response_fig}(c).  This temperature rise indicates that the thermal energy from the absorbed light diffuses through the YIG/GGG, but not enough to create a $\nabla T$ large enough to create an appreciable value of $V_{\rm ISHE}$.  

\subsection{Amplitude-Modulation Detection of the Spin Seebeck Effect}
\begin{figure}[h!]
	\includegraphics[width=1.00\columnwidth]{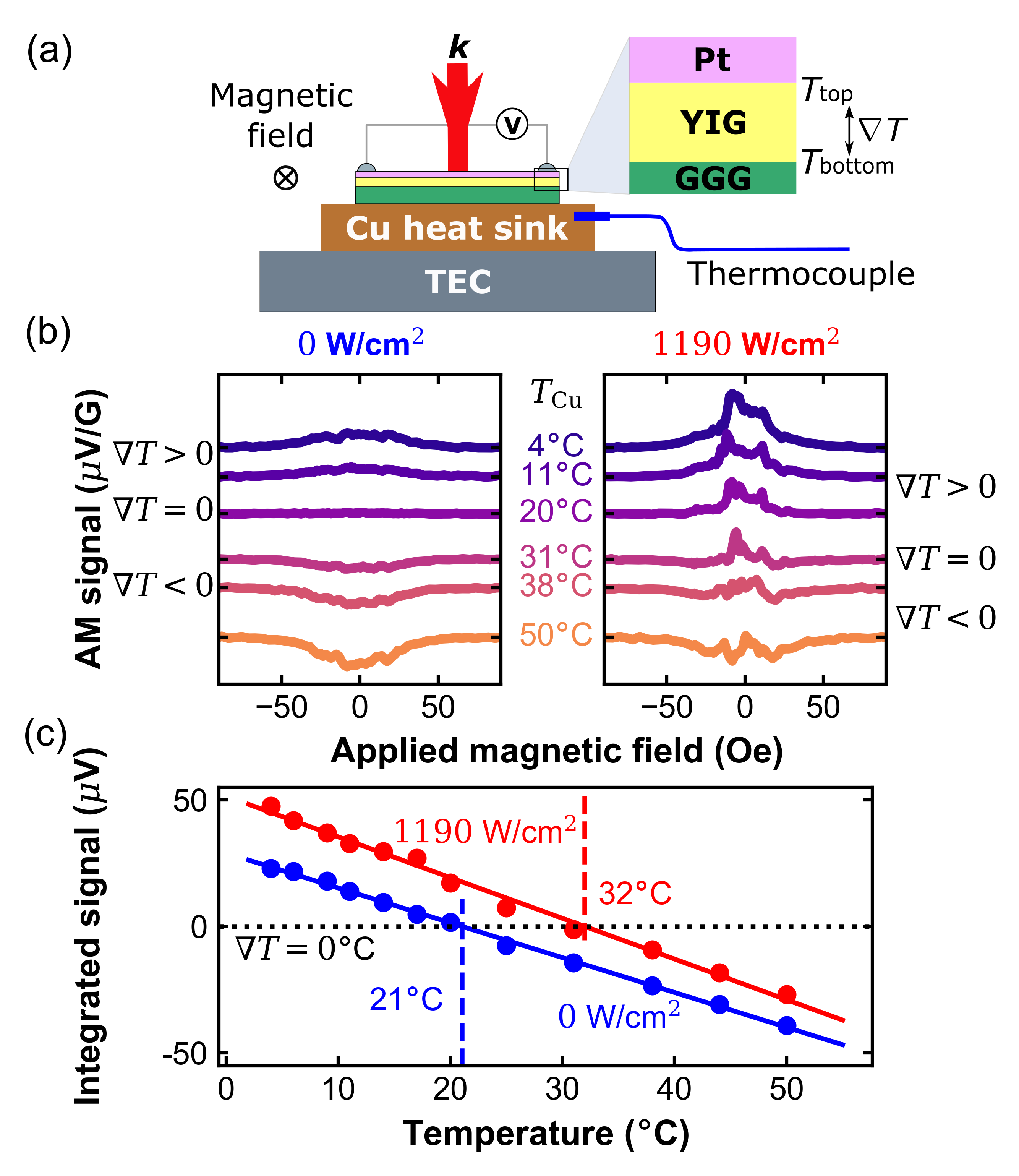}
	\caption{Measurement of the SSE using a field amplitude-modulation technique. 
	(a) Pt/YIG/GGG device mounted on a copper heat sink attached to a TEC for backside temperature control. (b) The amplitude-modulated (AM) signal with different copper heat sink temperatures, $T_{\rm Cu}$, with (right) and without (left) optical illumination.  As $T_{\rm Cu}$ is increased, the temperature gradient across the YIG, $\nabla T$, goes from negative to positive, which is mirrored by the sign of the AM signal. (c) Integrated AM signal with (red) and without (blue) optical illumination. The $\sim$11~K shift is created by the Pt absorption of the light and is directly attributable to the SSE.
	\label{Fig:figure5}}
\end{figure}

To resolve uncertainty in the determination of $\Delta T$, we modified our experiment to \textit{unambiguously} detect the SSE, as well as to precisely measure $\nabla T$.  In the reconfigured experimental setup, which is shown in Fig.~\ref{Fig:figure5}(a), the device is mounted on a copper heat sink, which is thermally mated to a thermoelectric cooler (TEC).  As before, a magnetic field is perpendicular to both the incoming light and the long axis of the device.  However, instead of modulating the optical intensity and sweeping the magnetic field, we removed the chopper and added a slow (13.1~Hz) and small ($\simeq$1~G) field modulation, $H_1$, to the larger, sweeping applied field, $H_0$.  By locking onto this modulation frequency, we obtain a voltage signal that corresponds to the derivative of the device response.  The key advantage of measuring this amplitude-modulated (AM) signal is that it allows us to study the SSE with or without light.  

The left panel in Fig.~\ref{Fig:figure5}(b) shows the AM signal as a function of the copper heat sink temperature, $T_{\rm Cu}$.  As discussed previously, the SSE occurs when $\nabla T$ is established across the YIG in the presence of $H_0$.  In our case, $\nabla T$ $\simeq\frac{T_{\rm Pt} - T_{\rm GGG}}{L_{\rm YIG}}$, where $T_{\rm Pt}$ and $T_{\rm GGG}$ are close approximations of the top- and bottom-side temperatures, respectively, of the YIG, and $L_{\rm YIG}$ is the YIG thickness.   For the light-off measurements, we assume that $T_{\rm Pt}$ is at room temperature ($\sim$20$^{\circ}$C).  Consequently, as we increase $T_{\rm Cu}$ ($\approx T_{\rm GGG}$) from 4 to 50$^{\circ}$C, $\nabla T$ correspondingly goes from positive to zero to negative, which is reflected in the magnitude and sign of the field-scanned AM signal.  The addition of a 1190 W/cm$^2$ intensity beam at 405 nm increases $T_{\rm Pt}$ and thus shifts $\nabla T$ by a fixed amount.  The right panel of Fig.~\ref{Fig:figure5}(b) shows exactly this behavior:  the point at which the AM signal is zero for the light-on condition occurs when $T_{\rm Cu}$ is at a larger value.  We note that more advanced modeling is necessary for an exact determination of $\nabla T$ under optical excitation, especially since the temperature across the Pt film is not constant~\cite{SurabhiScientificRep2018}.

Given that the AM signal is proportional to the derivative of the optical-modulation signal, the integration of the curves shown in Fig.~\ref{Fig:figure5}(b) is comparable to the SSE signal discussed in previous sections. Figure~\ref{Fig:figure5}(c) shows the integrated SSE AM signal as a function of $T_{\rm Cu}$ for both the light off (blue) and on (red). Since $V_{\rm ISHE} \propto J^{\rm SSE}_{\rm s} \propto \nabla T \simeq \left[T_{\rm Pt} - T_{\rm GGG}\right]\frac{1}{L_{\rm YIG}} \simeq \left[T_{\rm Pt} - T_{\rm Cu}\right]\frac{1}{L_{\rm YIG}}$, the key point here is that as the integrated AM signal is proportional to the magnitude and sign of the SSE-generated spin current density, $J^{\rm SSE}_{\rm s}$.  More importantly, the lateral difference between the light-on and -off curves is a highly accurate (relative) measure of thermal heating due to optical illumination:  the only thing changing between the curves shown in Fig.~\ref{Fig:figure5}(c) is the presence of an additional (optical) heating source.  This removes the inaccuracies incurred by thickness-dependent temperature gradients across the Pt film~\cite{SurabhiScientificRep2018} and sample-specific properties~\cite{AqeelJAP2014}.  Taking the point at which both curves cross zero ($\nabla T \simeq 0$), we find that when compared to the light-off condition, the optically created temperature difference between the top and bottom sides of the YIG is  $\sim$11~K using an intensity of 1190~W/cm$^2$.  This temperature difference due to the light, as well as the lack of heating due to optical absorption by the GGG substrate (Fig.~\ref{GGG_heating_fig}), suggests that we can accurately determine the optically created thermal gradient across the YIG to be: $\nabla T \simeq \frac{\Delta T}{L_{\rm YIG}} = 0.73$~K/$\mu$m.  Furthermore, we know that this intensity generates a $V_{\rm ISHE}$ of 3.457~$\mu$V across the $\sim$80~$\mu$m beam spot size, which is roughly measured by two indium contacts spaced 4.45~mm apart giving $E_{\rm ISHE} = -\nabla V_{\rm ISHE}$ = 3.457~$\mu$V/80~$\mu$m= 43~mV/m.  Using our estimates for $\nabla T$ and $-\nabla V_{\rm ISHE}$, we calculate a quantity akin to a traditional Seebeck coefficient for $S_{\rm LSSE}$ in this Pt/YIG device:  $S_{\rm LSSE} = -\nabla V_{\rm ISHE}/\nabla T = 60 \pm 7.8$ nV/K.  This estimate is in line with previous measurements of $S_{\rm LSSE}$ in Pt/YIG heterostructures~\cite{SolaJAP2015, SolaScientificReports2017, YuasaJPhysD2018, NakataJJAP2019}, but orders of magnitude below (electrical) Seebeck coefficients found in thermoelectric devices~\cite{BoukaiNature2008, YeJMaterChemC2017}.

Based off spectral, dynamic, and field-modulated measurements, we believe the observed $V_{\rm ISHE}$ is created via a spin current produced from the (optically heated) bulk SSE.  This interpretation does not exclude the possibility that the PSV effect~\cite{EllsworthNaturePhys2016} may also be present here.  However, a calculation using the estimated number of optically generated carriers in Pt for the powers we used in these experiments (details given in Appendix B), suggests that any PSV signal should be, at maximum, $\sim$4.9~nV.  Given that the measured $V_{\rm ISHE}$ is on the order of 10$^{-6}$~V, any PSV effect is likely negligible.  Moreover, the extremely flat, broadband, but low responsivity correlates well with (electrical) Seebeck effect devices, such as thermopiles.  Given the similarities, we expect a substantial increase in $\Re$ as temperature decreases, a hypothesis that preliminary measurements strongly support. \newline

\section{\label{sec:Conclusions}Conclusions}

Taken together, our results show a spin-based detection of broadband light across the Si-InGaAs detection range.  The spectral responsivity from 390 to 2200~nm is attributed to Pt absorption, which creates a spin current from the SSE.  We find that the thermal gradient across the underlying YIG layer from the incident light used to produce this spin current is 0.73~K/$\mu$m; this thermal gradient measurement allows us to estimate $S_{\rm LSSE}$ as $\sim$60 nV/K.  Unlike previous optical detectors that are fully reliant on charge carriers, the devices we examine here use spin current to produce a voltage response from light.  Our measurements of an ultrabroadband and featureless $\Re{}(\lambda)$, combined with previous SSE dynamical studies~\cite{AgrawalPRB2014, KimlingPRL2017} and routes for enhancing $S_{\rm LSSE}$~\cite{YuasaJPhysD2018, NakataJJAP2019}, suggest the promise of a spin-based optical detector that is competitive with current photovoltage architectures.  \newline

\section*{Acknowledgments} 
We thank Robert D.~McMichael and Mark D.~Stiles for insightful discussions.
The UW researchers acknowledge funding support from the Univ.~of WY School of Energy Resources. The CSU researchers acknowledge funding support from the U.~S.~National Science Foundation under Grants No.~EFMA-1641989 and No.~ECCS-1915849. \newline

\section*{Appendix A: Temperature change estimation using the resistance of platinum}

The resistance of the optically illuminated Pt was directly measured as a function of optical power to ascertain $\alpha^{\rm TC}_{\rm Pt}$, the resistance temperature coefficient.  Changes in temperature, $\Delta T$, are related to resistance changes via~\cite{ZhangJJAP1997}:
\begin{equation}
\Delta T = \frac{\Delta R}{R_0}\frac{1}{\alpha^{\rm TC}_{\rm Pt}},
\label{alpha_eqn}
\end{equation}
\noindent where $\Delta R$ is the change in the resistance and $R_0$ is the resistance under no optical illumination.
From the data shown in Fig.~\ref{Thermal_response_fig}(b), we determine $\alpha^{\rm TC}_{\rm Pt}$ to be 2.39$\times$10$^{-4}$~K$^{-1}$, which is significantly smaller than the $\alpha^{\rm TC}$ of bulk Pt, which is equal to 3.92$\times$10$^{-3}$~K$^{-1}$~\cite{Gale2004}. 



\section*{Appendix B: Magnitude of the photo-spin-voltaic effect}

Ellsworth et al.~\cite{EllsworthNaturePhys2016} have suggested that spin-polarized photogenerated carriers created in Pt and oriented via proximity to a magnetized insulator (in our case, YIG) can produce voltage signals similar to the ones we observe.  
In order to distinguish our results from the photo-spin-voltatic (PSV) effect, we estimate the magnitude from a 405~nm laser with an incident power of 18~mW and a beam diameter of 80~$\mu$m giving an intensity, $I$, of 3.6~MW/m$^2$.   
Starting with Eq.~\ref{ISHE_voltage_eqn}, we recognize that $D_{\rm ISHE}$ is equal to $-\theta_{\rm SH}\left(\frac{2e}{\hbar}\right)$, where $\theta_{\rm SH}$ is the spin Hall angle and has been reported to range from $10^{-4}$ to $10^0$ \cite{KimuraPRL2007, UchidaNature2008, SekiNatureMater2008, WangPRL2014}.  Since the charge current density, $\mathbf{J}_c$, equals $\sigma_c \mathbf{E}_{\rm ISHE}$, where $\sigma_c$ is measured to be $\sigma_c$ = 7.07$\times$$10^5$ S/m, we can estimate the magnitude of $\Delta V_{\rm ISHE}$ as:

\begin{equation}
\Delta V_{\rm ISHE} = \theta_{\rm SH} \left(\frac{2e\ell}{\hbar \sigma_c}\right) {J}_s,
\label{voltage_drop_eqn}
\end{equation}

\noindent where $\ell$ ($=$4.45$\times$$10^{-3}$ m) is the measured distance between the indium contacts.  

Next, we estimate the magnitude of $\mathbf{J}_s$ created by the incident light.  In the model put forth by Ellsworth and co-workers~\cite{EllsworthNaturePhys2016}, the electron spin density, $J^e_s$, is approximately zero, which means that the hole spin density, $J^h_s$, becomes the only contribution to $J_s$.  Thus, for the spin current density, we have~\cite{EllsworthNaturePhys2016}:  $J^h_s = J_s = -\frac{1}{2}E_{\rm light} d_0 t_{\rm Pt}(\epsilon''_{\uparrow}-\epsilon''_{\downarrow})$.  Here, $E_{\rm light}$ is the electric field of the light and $d_0$ is the region of Pt affected by the magnetization of the YIG, which we take to be 1 \AA~reflecting the small extent of the proximity effect.  It is only in this region that optically generated holes and electrons are spin polarized, which means that we need to only consider optical absorption within this thickness. As such, the light electric field at the ferromagnetic Pt region, $z_{\rm FM} = t_{\rm Pt}- d_0$, is: $E_{\rm light} (z_{\rm FM})= e^{-\frac{\alpha}{2}z_{\rm FM}}\sqrt{\frac{2 I}{c\epsilon_0}}$, where $\alpha^{\rm opt}_{\rm Pt}$ ($\simeq10^7$~m$^{-1}$) is the measured Pt absorption coefficient, $c$ is the speed of light, and $\epsilon_0$ is the vacuum permittivity.  We can therefore relate the light intensity to the magnitude of the hole spin density:

\begin{equation}
J_s = \frac{1}{2}E_{\rm light} (z_{\rm FM}) d_0 t_{\rm Pt}(\epsilon''_{\uparrow}-\epsilon''_{\downarrow})
\label{spin_density_eqn}
\end{equation}

\noindent Using Eqs.~\ref{voltage_drop_eqn} and \ref{spin_density_eqn}, the relationship between the magnitude of $\Delta V_{\rm ISHE}$ and $I$ for the PSV effect can thus be written as:

\begin{equation}
\Delta V_{\rm ISHE} = \theta_{\rm SH} \left(\frac{e\ell d_0 t_{\rm Pt}}{\hbar \sigma_c} \right)e^{-\frac{\alpha}{2}z_{\rm FM}}\sqrt{\frac{2 I}{c\epsilon_0}} (\epsilon''_{\uparrow}-\epsilon''_{\downarrow}).
\end{equation}

\noindent For our estimate, we take $\theta_{\rm SH} = 10^{-1}$~\cite{WangPRL2014} and $\epsilon''_{\uparrow}-\epsilon''_{\downarrow} \approx 0.5$~\cite{EllsworthNaturePhys2016}.  Plugging in these numbers, we can find that $\Delta V_{\rm ISHE}$ for our 2~nm-thick Pt device to be $\sim$4.9~nV, which is three orders of magnitude smaller than the observed signal, 3.457~$\mu$V, at this wavelength and optical intensity. Although the predicted wavelength dependence of $(\epsilon''_{\uparrow}-\epsilon''_{\downarrow})$ changes from positive to negative, its expected magnitude never goes above 5 (that is, a factor of ten from what we used in this estimate for 405 nm).  Increasing \textit{both} $\theta_{\rm SH}$ and $\epsilon''_{\uparrow}-\epsilon''_{\downarrow}$ by an order of magnitude still puts us a factor of ten below our empirical observations.  Thus, given the size of the estimated PSV effect signal for these intensities, as well as the lack of wavelength dependence observed in our measurements, we believe that the $V_{\rm ISHE}$ measured in our work comes predominantly from the bulk SSE. 


\begin{thebibliography}{52}%
\makeatletter
\providecommand \@ifxundefined [1]{%
 \@ifx{#1\undefined}
}%
\providecommand \@ifnum [1]{%
 \ifnum #1\expandafter \@firstoftwo
 \else \expandafter \@secondoftwo
 \fi
}%
\providecommand \@ifx [1]{%
 \ifx #1\expandafter \@firstoftwo
 \else \expandafter \@secondoftwo
 \fi
}%
\providecommand \natexlab [1]{#1}%
\providecommand \enquote  [1]{``#1''}%
\providecommand \bibnamefont  [1]{#1}%
\providecommand \bibfnamefont [1]{#1}%
\providecommand \citenamefont [1]{#1}%
\providecommand \href@noop [0]{\@secondoftwo}%
\providecommand \href [0]{\begingroup \@sanitize@url \@href}%
\providecommand \@href[1]{\@@startlink{#1}\@@href}%
\providecommand \@@href[1]{\endgroup#1\@@endlink}%
\providecommand \@sanitize@url [0]{\catcode `\\12\catcode `\$12\catcode
  `\&12\catcode `\#12\catcode `\^12\catcode `\_12\catcode `\%12\relax}%
\providecommand \@@startlink[1]{}%
\providecommand \@@endlink[0]{}%
\providecommand \url  [0]{\begingroup\@sanitize@url \@url }%
\providecommand \@url [1]{\endgroup\@href {#1}{\urlprefix }}%
\providecommand \urlprefix  [0]{URL }%
\providecommand \Eprint [0]{\href }%
\providecommand \doibase [0]{http://dx.doi.org/}%
\providecommand \selectlanguage [0]{\@gobble}%
\providecommand \bibinfo  [0]{\@secondoftwo}%
\providecommand \bibfield  [0]{\@secondoftwo}%
\providecommand \translation [1]{[#1]}%
\providecommand \BibitemOpen [0]{}%
\providecommand \bibitemStop [0]{}%
\providecommand \bibitemNoStop [0]{.\EOS\space}%
\providecommand \EOS [0]{\spacefactor3000\relax}%
\providecommand \BibitemShut  [1]{\csname bibitem#1\endcsname}%
\let\auto@bib@innerbib\@empty
\bibitem [{\citenamefont {Jungwirth}\ \emph {et~al.}(2012)\citenamefont
  {Jungwirth}, \citenamefont {Wunderlich},\ and\ \citenamefont
  {Olejn\'{i}k}}]{JungwirthNatureMater2012}%
  \BibitemOpen
  \bibfield  {author} {\bibinfo {author} {\bibfnamefont {T.}~\bibnamefont
  {Jungwirth}}, \bibinfo {author} {\bibfnamefont {J.}~\bibnamefont
  {Wunderlich}}, \ and\ \bibinfo {author} {\bibfnamefont {K.}~\bibnamefont
  {Olejn\'{i}k}},\ }\bibfield  {title} {\enquote {\bibinfo {title} {Spin Hall
  effect devices},}\ }\href@noop {} {\bibfield  {journal} {\bibinfo  {journal}
  {Nature Mater.}\ }\textbf {\bibinfo {volume} {11}},\ \bibinfo {pages} {382}
  (\bibinfo {year} {2012})}\BibitemShut {NoStop}%
\bibitem [{\citenamefont {Bauer}\ \emph {et~al.}(2012)\citenamefont {Bauer},
  \citenamefont {Saitoh},\ and\ \citenamefont {van
  Wees}}]{BauerNatureMater2012}%
  \BibitemOpen
  \bibfield  {author} {\bibinfo {author} {\bibfnamefont {G.~E.~W.}\
  \bibnamefont {Bauer}}, \bibinfo {author} {\bibfnamefont {E.}~\bibnamefont
  {Saitoh}}, \ and\ \bibinfo {author} {\bibfnamefont {B.~J.}\ \bibnamefont {van
  Wees}},\ }\bibfield  {title} {\enquote {\bibinfo {title} {Spin
  caloritronics},}\ }\href@noop {} {\bibfield  {journal} {\bibinfo  {journal}
  {Nature Mater.}\ }\textbf {\bibinfo {volume} {11}},\ \bibinfo {pages} {391}
  (\bibinfo {year} {2012})}\BibitemShut {NoStop}%
\bibitem [{\citenamefont {Wolf}\ \emph {et~al.}(2001)\citenamefont {Wolf},
  \citenamefont {Awschalom}, \citenamefont {Buhrman}, \citenamefont {Daughton},
  \citenamefont {von Molnar}, \citenamefont {Roukes}, \citenamefont
  {Chtchelkanova},\ and\ \citenamefont {Treger}}]{WolfScience2001}%
  \BibitemOpen
  \bibfield  {author} {\bibinfo {author} {\bibfnamefont {S.~A.}\ \bibnamefont
  {Wolf}}, \bibinfo {author} {\bibfnamefont {D.~D.}\ \bibnamefont {Awschalom}},
  \bibinfo {author} {\bibfnamefont {R.~A.}\ \bibnamefont {Buhrman}}, \bibinfo
  {author} {\bibfnamefont {J.~M.}\ \bibnamefont {Daughton}}, \bibinfo {author}
  {\bibfnamefont {S.}~\bibnamefont {von Molnar}}, \bibinfo {author}
  {\bibfnamefont {M.~L.}\ \bibnamefont {Roukes}}, \bibinfo {author}
  {\bibfnamefont {A.~Y.}\ \bibnamefont {Chtchelkanova}}, \ and\ \bibinfo
  {author} {\bibfnamefont {D.~M.}\ \bibnamefont {Treger}},\ }\bibfield  {title}
  {\enquote {\bibinfo {title} {Spintronics: A spin-based electronics vision for
  the future},}\ }\href@noop {} {\bibfield  {journal} {\bibinfo  {journal}
  {Science}\ }\textbf {\bibinfo {volume} {294}},\ \bibinfo {pages} {1488--1495}
  (\bibinfo {year} {2001})}\BibitemShut {NoStop}%
\bibitem [{\citenamefont {\u{Z}uti\'{c}}\ \emph {et~al.}(2004)\citenamefont
  {\u{Z}uti\'{c}}, \citenamefont {Fabian},\ and\ \citenamefont
  {Sarma}}]{ZuticRMP2004}%
  \BibitemOpen
  \bibfield  {author} {\bibinfo {author} {\bibfnamefont {I.}~\bibnamefont
  {\u{Z}uti\'{c}}}, \bibinfo {author} {\bibfnamefont {J.}~\bibnamefont
  {Fabian}}, \ and\ \bibinfo {author} {\bibfnamefont {S.~Das}\ \bibnamefont
  {Sarma}},\ }\bibfield  {title} {\enquote {\bibinfo {title} {Spintronics:
  Fundamentals and applications},}\ }\href@noop {} {\bibfield  {journal}
  {\bibinfo  {journal} {Rev. Mod. Phys.}\ }\textbf {\bibinfo {volume} {76}},\
  \bibinfo {pages} {323} (\bibinfo {year} {2004})}\BibitemShut {NoStop}%
\bibitem [{\citenamefont {N\'{a}fr\'{a}di}\ \emph {et~al.}(2016)\citenamefont
  {N\'{a}fr\'{a}di}, \citenamefont {Szirmai}, \citenamefont {Spina},
  \citenamefont {Lee}, \citenamefont {Yazyev}, \citenamefont {Arakcheeva},
  \citenamefont {Chernyshov}, \citenamefont {Gibert}, \citenamefont
  {Forr\'{o}},\ and\ \citenamefont {Horv\'{a}th}}]{NafradiNatureComm2016}%
  \BibitemOpen
  \bibfield  {author} {\bibinfo {author} {\bibfnamefont {B.}~\bibnamefont
  {N\'{a}fr\'{a}di}}, \bibinfo {author} {\bibfnamefont {P.}~\bibnamefont
  {Szirmai}}, \bibinfo {author} {\bibfnamefont {M.}~\bibnamefont {Spina}},
  \bibinfo {author} {\bibfnamefont {H.}~\bibnamefont {Lee}}, \bibinfo {author}
  {\bibfnamefont {O.~V.}\ \bibnamefont {Yazyev}}, \bibinfo {author}
  {\bibfnamefont {A.}~\bibnamefont {Arakcheeva}}, \bibinfo {author}
  {\bibfnamefont {D.}~\bibnamefont {Chernyshov}}, \bibinfo {author}
  {\bibfnamefont {M.}~\bibnamefont {Gibert}}, \bibinfo {author} {\bibfnamefont
  {L.}~\bibnamefont {Forr\'{o}}}, \ and\ \bibinfo {author} {\bibfnamefont
  {E.}~\bibnamefont {Horv\'{a}th}},\ }\bibfield  {title} {\enquote {\bibinfo
  {title} {Optically switched magnetism in photovoltaic perovskite
  CH$_3$NH$_3$(Mn:Pb)I$_3$},}\ }\href@noop {} {\bibfield  {journal} {\bibinfo
  {journal} {Nature Comm.}\ }\textbf {\bibinfo {volume} {7}},\ \bibinfo {pages}
  {13406} (\bibinfo {year} {2016})}\BibitemShut {NoStop}%
\bibitem [{\citenamefont {Uchida}\ \emph {et~al.}(2008)\citenamefont {Uchida},
  \citenamefont {Takahashi}, \citenamefont {Harii}, \citenamefont {Ieda},
  \citenamefont {Koshibae}, \citenamefont {Ando}, \citenamefont {Maekawa},\
  and\ \citenamefont {Saitoh}}]{UchidaNature2008}%
  \BibitemOpen
  \bibfield  {author} {\bibinfo {author} {\bibfnamefont {K.}~\bibnamefont
  {Uchida}}, \bibinfo {author} {\bibfnamefont {S.}~\bibnamefont {Takahashi}},
  \bibinfo {author} {\bibfnamefont {K.}~\bibnamefont {Harii}}, \bibinfo
  {author} {\bibfnamefont {J.}~\bibnamefont {Ieda}}, \bibinfo {author}
  {\bibfnamefont {W.}~\bibnamefont {Koshibae}}, \bibinfo {author}
  {\bibfnamefont {K.}~\bibnamefont {Ando}}, \bibinfo {author} {\bibfnamefont
  {S.}~\bibnamefont {Maekawa}}, \ and\ \bibinfo {author} {\bibfnamefont
  {E.}~\bibnamefont {Saitoh}},\ }\bibfield  {title} {\enquote {\bibinfo {title}
  {Observation of the spin Seebeck effect},}\ }\href@noop {} {\bibfield
  {journal} {\bibinfo  {journal} {Nature}\ }\textbf {\bibinfo {volume} {455}},\
  \bibinfo {pages} {778} (\bibinfo {year} {2008})}\BibitemShut {NoStop}%
\bibitem [{\citenamefont {Azevedo}\ \emph {et~al.}(2005)\citenamefont
  {Azevedo}, \citenamefont {{Vilela Le\~{a}o}}, \citenamefont
  {Rodriguez-Suarez}, \citenamefont {Oliveira},\ and\ \citenamefont
  {Rezende}}]{AzevedoJAP2005}%
  \BibitemOpen
  \bibfield  {author} {\bibinfo {author} {\bibfnamefont {A.}~\bibnamefont
  {Azevedo}}, \bibinfo {author} {\bibfnamefont {L.~H.}\ \bibnamefont {{Vilela
  Le\~{a}o}}}, \bibinfo {author} {\bibfnamefont {R.~L}\ \bibnamefont
  {Rodriguez-Suarez}}, \bibinfo {author} {\bibfnamefont {A.~B.}\ \bibnamefont
  {Oliveira}}, \ and\ \bibinfo {author} {\bibfnamefont {S.~M.}\ \bibnamefont
  {Rezende}},\ }\bibfield  {title} {\enquote {\bibinfo {title} {dc effect in
  ferromagnetic resonance: Evidence of the spin-pumping effect?}}\ }\href@noop
  {} {\bibfield  {journal} {\bibinfo  {journal} {J. Appl. Phys.}\ }\textbf
  {\bibinfo {volume} {97}},\ \bibinfo {pages} {10C715} (\bibinfo {year}
  {2005})}\BibitemShut {NoStop}%
\bibitem [{\citenamefont {Saitoh}\ \emph {et~al.}(2006)\citenamefont {Saitoh},
  \citenamefont {Ueda}, \citenamefont {Miyajima},\ and\ \citenamefont
  {Tatara}}]{SaitohAPL2006}%
  \BibitemOpen
  \bibfield  {author} {\bibinfo {author} {\bibfnamefont {E.}~\bibnamefont
  {Saitoh}}, \bibinfo {author} {\bibfnamefont {M.}~\bibnamefont {Ueda}},
  \bibinfo {author} {\bibfnamefont {H.}~\bibnamefont {Miyajima}}, \ and\
  \bibinfo {author} {\bibfnamefont {G.}~\bibnamefont {Tatara}},\ }\bibfield
  {title} {\enquote {\bibinfo {title} {Conversion of spin current into charge
  current at room temperature: Inverse spin-Hall effect},}\ }\href@noop {}
  {\bibfield  {journal} {\bibinfo  {journal} {Appl. Phys. Lett.}\ }\textbf
  {\bibinfo {volume} {88}},\ \bibinfo {pages} {182509} (\bibinfo {year}
  {2006})}\BibitemShut {NoStop}%
\bibitem [{\citenamefont {Hirsch}(1999)}]{HirschPRL1999}%
  \BibitemOpen
  \bibfield  {author} {\bibinfo {author} {\bibfnamefont {J.~E.}\ \bibnamefont
  {Hirsch}},\ }\bibfield  {title} {\enquote {\bibinfo {title} {Spin Hall
  effect},}\ }\href@noop {} {\bibfield  {journal} {\bibinfo  {journal} {Phys.
  Rev. Lett.}\ }\textbf {\bibinfo {volume} {83}},\ \bibinfo {pages} {1834}
  (\bibinfo {year} {1999})}\BibitemShut {NoStop}%
\bibitem [{\citenamefont {Kato}\ \emph {et~al.}(2004)\citenamefont {Kato},
  \citenamefont {Myers}, \citenamefont {Gossard},\ and\ \citenamefont
  {Awschalom}}]{KatoScience2004}%
  \BibitemOpen
  \bibfield  {author} {\bibinfo {author} {\bibfnamefont {Y.~K.}\ \bibnamefont
  {Kato}}, \bibinfo {author} {\bibfnamefont {R.~C.}\ \bibnamefont {Myers}},
  \bibinfo {author} {\bibfnamefont {A.~C.}\ \bibnamefont {Gossard}}, \ and\
  \bibinfo {author} {\bibfnamefont {D.~D.}\ \bibnamefont {Awschalom}},\
  }\bibfield  {title} {\enquote {\bibinfo {title} {Observation of the spin Hall
  effect in semiconductors},}\ }\href@noop {} {\bibfield  {journal} {\bibinfo
  {journal} {Science}\ }\textbf {\bibinfo {volume} {306}},\ \bibinfo {pages}
  {1910} (\bibinfo {year} {2004})}\BibitemShut {NoStop}%
\bibitem [{\citenamefont {Valenzuela}\ and\ \citenamefont
  {Tinkham}(2006)}]{ValenzuelaNature2006}%
  \BibitemOpen
  \bibfield  {author} {\bibinfo {author} {\bibfnamefont {S.~O.}\ \bibnamefont
  {Valenzuela}}\ and\ \bibinfo {author} {\bibfnamefont {M.}~\bibnamefont
  {Tinkham}},\ }\bibfield  {title} {\enquote {\bibinfo {title} {Direct
  electronic measurement of the spin Hall effect},}\ }\href@noop {} {\bibfield
  {journal} {\bibinfo  {journal} {Nature}\ }\textbf {\bibinfo {volume} {442}},\
  \bibinfo {pages} {176} (\bibinfo {year} {2006})}\BibitemShut {NoStop}%
\bibitem [{\citenamefont {Kirihara}\ \emph {et~al.}(2012)\citenamefont
  {Kirihara}, \citenamefont {Uchida}, \citenamefont {Kajiwara}, \citenamefont
  {Ishida}, \citenamefont {Nakamura}, \citenamefont {Manako}, \citenamefont
  {Saitoh},\ and\ \citenamefont {Yorozu}}]{KiriharaNatureMater2012}%
  \BibitemOpen
  \bibfield  {author} {\bibinfo {author} {\bibfnamefont {A.}~\bibnamefont
  {Kirihara}}, \bibinfo {author} {\bibfnamefont {K.}~\bibnamefont {Uchida}},
  \bibinfo {author} {\bibfnamefont {Y.}~\bibnamefont {Kajiwara}}, \bibinfo
  {author} {\bibfnamefont {M.}~\bibnamefont {Ishida}}, \bibinfo {author}
  {\bibfnamefont {Y.}~\bibnamefont {Nakamura}}, \bibinfo {author}
  {\bibfnamefont {T.}~\bibnamefont {Manako}}, \bibinfo {author} {\bibfnamefont
  {E.}~\bibnamefont {Saitoh}}, \ and\ \bibinfo {author} {\bibfnamefont
  {S.}~\bibnamefont {Yorozu}},\ }\bibfield  {title} {\enquote {\bibinfo {title}
  {Spin-current-driven thermoelectric coating},}\ }\href@noop {} {\bibfield
  {journal} {\bibinfo  {journal} {Nature Mater.}\ }\textbf {\bibinfo {volume}
  {11}},\ \bibinfo {pages} {686} (\bibinfo {year} {2012})}\BibitemShut
  {NoStop}%
\bibitem [{\citenamefont {Uchida}\ \emph {et~al.}(2012)\citenamefont {Uchida},
  \citenamefont {Ota}, \citenamefont {Adachi}, \citenamefont {Xiao},
  \citenamefont {Nonaka}, \citenamefont {Kajiwara}, \citenamefont {Bauer},
  \citenamefont {Maekawa},\ and\ \citenamefont {Saitoh}}]{UchidaJAP2012}%
  \BibitemOpen
  \bibfield  {author} {\bibinfo {author} {\bibfnamefont {K.}~\bibnamefont
  {Uchida}}, \bibinfo {author} {\bibfnamefont {T.}~\bibnamefont {Ota}},
  \bibinfo {author} {\bibfnamefont {H.}~\bibnamefont {Adachi}}, \bibinfo
  {author} {\bibfnamefont {J.}~\bibnamefont {Xiao}}, \bibinfo {author}
  {\bibfnamefont {T.}~\bibnamefont {Nonaka}}, \bibinfo {author} {\bibfnamefont
  {Y.}~\bibnamefont {Kajiwara}}, \bibinfo {author} {\bibfnamefont {G.~E.~W.}\
  \bibnamefont {Bauer}}, \bibinfo {author} {\bibfnamefont {S.}~\bibnamefont
  {Maekawa}}, \ and\ \bibinfo {author} {\bibfnamefont {E.}~\bibnamefont
  {Saitoh}},\ }\bibfield  {title} {\enquote {\bibinfo {title} {Thermal spin
  pumping and magnon-phonon-mediated spin-Seebeck effect},}\ }\href@noop {}
  {\bibfield  {journal} {\bibinfo  {journal} {J. Appl. Phys.}\ }\textbf
  {\bibinfo {volume} {111}},\ \bibinfo {pages} {103903} (\bibinfo {year}
  {2012})}\BibitemShut {NoStop}%
\bibitem [{\citenamefont {Jaworski}\ \emph {et~al.}(2010)\citenamefont
  {Jaworski}, \citenamefont {Yang}, \citenamefont {Mack}, \citenamefont
  {Awschalom}, \citenamefont {Heremans},\ and\ \citenamefont
  {Myers}}]{JaworskiNatureMater2010}%
  \BibitemOpen
  \bibfield  {author} {\bibinfo {author} {\bibfnamefont {C.~M.}\ \bibnamefont
  {Jaworski}}, \bibinfo {author} {\bibfnamefont {J.}~\bibnamefont {Yang}},
  \bibinfo {author} {\bibfnamefont {S.}~\bibnamefont {Mack}}, \bibinfo {author}
  {\bibfnamefont {D.~D.}\ \bibnamefont {Awschalom}}, \bibinfo {author}
  {\bibfnamefont {J.~P.}\ \bibnamefont {Heremans}}, \ and\ \bibinfo {author}
  {\bibfnamefont {R.~C.}\ \bibnamefont {Myers}},\ }\bibfield  {title} {\enquote
  {\bibinfo {title} {Observation of the spin-Seebeck effect in a ferromagnetic
  semiconductor},}\ }\href@noop {} {\bibfield  {journal} {\bibinfo  {journal}
  {Nature Mater.}\ }\textbf {\bibinfo {volume} {9}},\ \bibinfo {pages} {898}
  (\bibinfo {year} {2010})}\BibitemShut {NoStop}%
\bibitem [{\citenamefont {Uchida}\ \emph
  {et~al.}(2010{\natexlab{a}})\citenamefont {Uchida}, \citenamefont {Adachi},
  \citenamefont {Ota}, \citenamefont {Nakayama}, \citenamefont {Maekawa},\ and\
  \citenamefont {Saitoh}}]{UchidaAPL2010}%
  \BibitemOpen
  \bibfield  {author} {\bibinfo {author} {\bibfnamefont {K.}~\bibnamefont
  {Uchida}}, \bibinfo {author} {\bibfnamefont {H.}~\bibnamefont {Adachi}},
  \bibinfo {author} {\bibfnamefont {T.}~\bibnamefont {Ota}}, \bibinfo {author}
  {\bibfnamefont {H.}~\bibnamefont {Nakayama}}, \bibinfo {author}
  {\bibfnamefont {S.}~\bibnamefont {Maekawa}}, \ and\ \bibinfo {author}
  {\bibfnamefont {E.}~\bibnamefont {Saitoh}},\ }\bibfield  {title} {\enquote
  {\bibinfo {title} {Observation of longitudinal spin-Seebeck effect in
  magnetic insulators},}\ }\href@noop {} {\bibfield  {journal} {\bibinfo
  {journal} {Appl. Phys. Lett.}\ }\textbf {\bibinfo {volume} {97}},\ \bibinfo
  {pages} {172505} (\bibinfo {year} {2010}{\natexlab{a}})}\BibitemShut
  {NoStop}%
\bibitem [{\citenamefont {Uchida}\ \emph
  {et~al.}(2010{\natexlab{b}})\citenamefont {Uchida}, \citenamefont {Nonaka},
  \citenamefont {Ota},\ and\ \citenamefont {Saitoh}}]{UchidaAPL2010_2}%
  \BibitemOpen
  \bibfield  {author} {\bibinfo {author} {\bibfnamefont {K.}~\bibnamefont
  {Uchida}}, \bibinfo {author} {\bibfnamefont {T.}~\bibnamefont {Nonaka}},
  \bibinfo {author} {\bibfnamefont {T.}~\bibnamefont {Ota}}, \ and\ \bibinfo
  {author} {\bibfnamefont {E.}~\bibnamefont {Saitoh}},\ }\bibfield  {title}
  {\enquote {\bibinfo {title} {Longitudinal spin-Seebeck effect in sintered
  polycrystalline (Mn, Zn)Fe$_2$O$_4$},}\ }\href@noop {} {\bibfield  {journal}
  {\bibinfo  {journal} {Appl. Phys. Lett.}\ }\textbf {\bibinfo {volume} {97}},\
  \bibinfo {pages} {262504} (\bibinfo {year} {2010}{\natexlab{b}})}\BibitemShut
  {NoStop}%
\bibitem [{\citenamefont {Uchida}\ \emph
  {et~al.}(2010{\natexlab{c}})\citenamefont {Uchida}, \citenamefont {Xiao},
  \citenamefont {Adachi}, \citenamefont {Ohe}, \citenamefont {Takahashi},
  \citenamefont {Ieda}, \citenamefont {Ota}, \citenamefont {Kajiwara},
  \citenamefont {Umezawa}, \citenamefont {Kawai}, \citenamefont {Bauer},
  \citenamefont {Maekawa},\ and\ \citenamefont
  {Saitoh}}]{UchidaNatureMater2010}%
  \BibitemOpen
  \bibfield  {author} {\bibinfo {author} {\bibfnamefont {K.}~\bibnamefont
  {Uchida}}, \bibinfo {author} {\bibfnamefont {J.}~\bibnamefont {Xiao}},
  \bibinfo {author} {\bibfnamefont {H.}~\bibnamefont {Adachi}}, \bibinfo
  {author} {\bibfnamefont {J.}~\bibnamefont {Ohe}}, \bibinfo {author}
  {\bibfnamefont {S.}~\bibnamefont {Takahashi}}, \bibinfo {author}
  {\bibfnamefont {J.}~\bibnamefont {Ieda}}, \bibinfo {author} {\bibfnamefont
  {T.}~\bibnamefont {Ota}}, \bibinfo {author} {\bibfnamefont {Y.}~\bibnamefont
  {Kajiwara}}, \bibinfo {author} {\bibfnamefont {H.}~\bibnamefont {Umezawa}},
  \bibinfo {author} {\bibfnamefont {H.}~\bibnamefont {Kawai}}, \bibinfo
  {author} {\bibfnamefont {G.~E.~W.}\ \bibnamefont {Bauer}}, \bibinfo {author}
  {\bibfnamefont {S.}~\bibnamefont {Maekawa}}, \ and\ \bibinfo {author}
  {\bibfnamefont {E.}~\bibnamefont {Saitoh}},\ }\bibfield  {title} {\enquote
  {\bibinfo {title} {Spin Seebeck insulator},}\ }\href@noop {} {\bibfield
  {journal} {\bibinfo  {journal} {Nature Mater.}\ }\textbf {\bibinfo {volume}
  {9}},\ \bibinfo {pages} {894} (\bibinfo {year}
  {2010}{\natexlab{c}})}\BibitemShut {NoStop}%
\bibitem [{\citenamefont {Ellsworth}\ \emph {et~al.}(2016)\citenamefont
  {Ellsworth}, \citenamefont {Lei}, \citenamefont {Lan}, \citenamefont {Chang},
  \citenamefont {Li}, \citenamefont {Wang}, \citenamefont {Hu}, \citenamefont
  {Johnson}, \citenamefont {Bian}, \citenamefont {Xiao}, \citenamefont {Wu},\
  and\ \citenamefont {Wu}}]{EllsworthNaturePhys2016}%
  \BibitemOpen
  \bibfield  {author} {\bibinfo {author} {\bibfnamefont {D.}~\bibnamefont
  {Ellsworth}}, \bibinfo {author} {\bibfnamefont {L.}~\bibnamefont {Lei}},
  \bibinfo {author} {\bibfnamefont {J.}~\bibnamefont {Lan}}, \bibinfo {author}
  {\bibfnamefont {H.}~\bibnamefont {Chang}}, \bibinfo {author} {\bibfnamefont
  {P.}~\bibnamefont {Li}}, \bibinfo {author} {\bibfnamefont {Z.}~\bibnamefont
  {Wang}}, \bibinfo {author} {\bibfnamefont {J.}~\bibnamefont {Hu}}, \bibinfo
  {author} {\bibfnamefont {B.}~\bibnamefont {Johnson}}, \bibinfo {author}
  {\bibfnamefont {Y.}~\bibnamefont {Bian}}, \bibinfo {author} {\bibfnamefont
  {J.}~\bibnamefont {Xiao}}, \bibinfo {author} {\bibfnamefont {R.}~\bibnamefont
  {Wu}}, \ and\ \bibinfo {author} {\bibfnamefont {M.}~\bibnamefont {Wu}},\
  }\bibfield  {title} {\enquote {\bibinfo {title} {Photo-spin-voltaic
  effect},}\ }\href@noop {} {\bibfield  {journal} {\bibinfo  {journal} {Nature
  Phys.}\ }\textbf {\bibinfo {volume} {12}},\ \bibinfo {pages} {861} (\bibinfo
  {year} {2016})}\BibitemShut {NoStop}%
\bibitem [{\citenamefont {\u{Z}uti\'{c}}\ and\ \citenamefont
  {Fabian}(2003)}]{ZuticMaterTrans2003}%
  \BibitemOpen
  \bibfield  {author} {\bibinfo {author} {\bibfnamefont {I.}~\bibnamefont
  {\u{Z}uti\'{c}}}\ and\ \bibinfo {author} {\bibfnamefont {J.}~\bibnamefont
  {Fabian}},\ }\bibfield  {title} {\enquote {\bibinfo {title} {Spin-voltaic
  effect and its implications},}\ }\href@noop {} {\bibfield  {journal}
  {\bibinfo  {journal} {Materials Trans.}\ }\textbf {\bibinfo {volume} {44}},\
  \bibinfo {pages} {2062} (\bibinfo {year} {2003})}\BibitemShut {NoStop}%
\bibitem [{\citenamefont {Kimura}\ \emph {et~al.}(2007)\citenamefont {Kimura},
  \citenamefont {Otani}, \citenamefont {Sato}, \citenamefont {Takahashi},\ and\
  \citenamefont {Maekawa}}]{KimuraPRL2007}%
  \BibitemOpen
  \bibfield  {author} {\bibinfo {author} {\bibfnamefont {T.}~\bibnamefont
  {Kimura}}, \bibinfo {author} {\bibfnamefont {Y.}~\bibnamefont {Otani}},
  \bibinfo {author} {\bibfnamefont {T.}~\bibnamefont {Sato}}, \bibinfo {author}
  {\bibfnamefont {S.}~\bibnamefont {Takahashi}}, \ and\ \bibinfo {author}
  {\bibfnamefont {S.}~\bibnamefont {Maekawa}},\ }\bibfield  {title} {\enquote
  {\bibinfo {title} {Room-temperature reversible spin Hall effect},}\
  }\href@noop {} {\bibfield  {journal} {\bibinfo  {journal} {Phys. Rev. Lett.}\
  }\textbf {\bibinfo {volume} {98}},\ \bibinfo {pages} {156601} (\bibinfo
  {year} {2007})}\BibitemShut {NoStop}%
\bibitem [{\citenamefont {Bartell}\ \emph {et~al.}(2017)\citenamefont
  {Bartell}, \citenamefont {Jermain}, \citenamefont {Aradhya}, \citenamefont
  {Brangham}, \citenamefont {Yang}, \citenamefont {Ralph},\ and\ \citenamefont
  {Fuchs}}]{BartellPRAppl2017}%
  \BibitemOpen
  \bibfield  {author} {\bibinfo {author} {\bibfnamefont {J.~M.}\ \bibnamefont
  {Bartell}}, \bibinfo {author} {\bibfnamefont {C.~L.}\ \bibnamefont
  {Jermain}}, \bibinfo {author} {\bibfnamefont {S.~V.}\ \bibnamefont
  {Aradhya}}, \bibinfo {author} {\bibfnamefont {J.~T.}\ \bibnamefont
  {Brangham}}, \bibinfo {author} {\bibfnamefont {F.}~\bibnamefont {Yang}},
  \bibinfo {author} {\bibfnamefont {D.~C.}\ \bibnamefont {Ralph}}, \ and\
  \bibinfo {author} {\bibfnamefont {G.~D.}\ \bibnamefont {Fuchs}},\ }\bibfield
  {title} {\enquote {\bibinfo {title} {Imaging magnetization structure and
  dynamics in ultrathin Y$_3$Fe$_5$O$_{12}$/Pt bilayers with high sensitivity
  using the time-resolved longitudinal spin Seebeck effect},}\ }\href@noop {}
  {\bibfield  {journal} {\bibinfo  {journal} {Phys. Rev. Appl.}\ }\textbf
  {\bibinfo {volume} {7}},\ \bibinfo {pages} {044004} (\bibinfo {year}
  {2017})}\BibitemShut {NoStop}%
\bibitem [{\citenamefont {Choi}\ \emph {et~al.}(2014)\citenamefont {Choi},
  \citenamefont {Min}, \citenamefont {Lee},\ and\ \citenamefont
  {Cahill}}]{ChoiNatureComm2014}%
  \BibitemOpen
  \bibfield  {author} {\bibinfo {author} {\bibfnamefont {G-M.}\ \bibnamefont
  {Choi}}, \bibinfo {author} {\bibfnamefont {B-C.}\ \bibnamefont {Min}},
  \bibinfo {author} {\bibfnamefont {K-J.}\ \bibnamefont {Lee}}, \ and\ \bibinfo
  {author} {\bibfnamefont {D.~G.}\ \bibnamefont {Cahill}},\ }\bibfield  {title}
  {\enquote {\bibinfo {title} {Spin current generated by thermally driven
  ultrafast demagnetization},}\ }\href@noop {} {\bibfield  {journal} {\bibinfo
  {journal} {Nature Comm.}\ }\textbf {\bibinfo {volume} {5}},\ \bibinfo {pages}
  {4334} (\bibinfo {year} {2014})}\BibitemShut {NoStop}%
\bibitem [{\citenamefont {Kimling}\ \emph {et~al.}(2017)\citenamefont
  {Kimling}, \citenamefont {Choi}, \citenamefont {Brangham}, \citenamefont
  {Matalla-Wagner}, \citenamefont {Huebner}, \citenamefont {Kuschel},
  \citenamefont {Yang},\ and\ \citenamefont {Cahill}}]{KimlingPRL2017}%
  \BibitemOpen
  \bibfield  {author} {\bibinfo {author} {\bibfnamefont {J.}~\bibnamefont
  {Kimling}}, \bibinfo {author} {\bibfnamefont {G.-M.}\ \bibnamefont {Choi}},
  \bibinfo {author} {\bibfnamefont {J.~T.}\ \bibnamefont {Brangham}}, \bibinfo
  {author} {\bibfnamefont {T.}~\bibnamefont {Matalla-Wagner}}, \bibinfo
  {author} {\bibfnamefont {T.}~\bibnamefont {Huebner}}, \bibinfo {author}
  {\bibfnamefont {T.}~\bibnamefont {Kuschel}}, \bibinfo {author} {\bibfnamefont
  {F.}~\bibnamefont {Yang}}, \ and\ \bibinfo {author} {\bibfnamefont {D.~G.}\
  \bibnamefont {Cahill}},\ }\bibfield  {title} {\enquote {\bibinfo {title}
  {Picosecond spin Seebeck effect},}\ }\href@noop {} {\bibfield  {journal}
  {\bibinfo  {journal} {Phys. Rev. Lett.}\ }\textbf {\bibinfo {volume} {118}},\
  \bibinfo {pages} {057201} (\bibinfo {year} {2017})}\BibitemShut {NoStop}%
\bibitem [{\citenamefont {McLaughlin}\ \emph {et~al.}(2017)\citenamefont
  {McLaughlin}, \citenamefont {Sun}, \citenamefont {Zhang}, \citenamefont
  {Groesbeck},\ and\ \citenamefont {Vardeny}}]{McLaughlinPRB2017}%
  \BibitemOpen
  \bibfield  {author} {\bibinfo {author} {\bibfnamefont {R.}~\bibnamefont
  {McLaughlin}}, \bibinfo {author} {\bibfnamefont {D.}~\bibnamefont {Sun}},
  \bibinfo {author} {\bibfnamefont {C.}~\bibnamefont {Zhang}}, \bibinfo
  {author} {\bibfnamefont {M.}~\bibnamefont {Groesbeck}}, \ and\ \bibinfo
  {author} {\bibfnamefont {Z.~V.}\ \bibnamefont {Vardeny}},\ }\bibfield
  {title} {\enquote {\bibinfo {title} {Optical detection of transverse
  spin-Seebeck effect in permalloy film using Sagnac interferometer
  microscopy},}\ }\href@noop {} {\bibfield  {journal} {\bibinfo  {journal}
  {Phys. Rev. B}\ }\textbf {\bibinfo {volume} {95}},\ \bibinfo {pages}
  {180401(R)} (\bibinfo {year} {2017})}\BibitemShut {NoStop}%
\bibitem [{\citenamefont {Sakai}\ \emph {et~al.}(2016)\citenamefont {Sakai},
  \citenamefont {Mizuta}, \citenamefont {Nugroho}, \citenamefont {Sihombing},
  \citenamefont {Koretsune}, \citenamefont {Suzuki}, \citenamefont {Takemori},
  \citenamefont {Ishii}, \citenamefont {Nishio-Hamane}, \citenamefont {Arita},
  \citenamefont {Goswami},\ and\ \citenamefont
  {Nakatsuji}}]{SakaiNaturePhys2016}%
  \BibitemOpen
  \bibfield  {author} {\bibinfo {author} {\bibfnamefont {A.}~\bibnamefont
  {Sakai}}, \bibinfo {author} {\bibfnamefont {Y.~P.}\ \bibnamefont {Mizuta}},
  \bibinfo {author} {\bibfnamefont {A.~A.}\ \bibnamefont {Nugroho}}, \bibinfo
  {author} {\bibfnamefont {R.}~\bibnamefont {Sihombing}}, \bibinfo {author}
  {\bibfnamefont {T.}~\bibnamefont {Koretsune}}, \bibinfo {author}
  {\bibfnamefont {M.-T.}\ \bibnamefont {Suzuki}}, \bibinfo {author}
  {\bibfnamefont {N.}~\bibnamefont {Takemori}}, \bibinfo {author}
  {\bibfnamefont {R.}~\bibnamefont {Ishii}}, \bibinfo {author} {\bibfnamefont
  {D.}~\bibnamefont {Nishio-Hamane}}, \bibinfo {author} {\bibfnamefont
  {R.}~\bibnamefont {Arita}}, \bibinfo {author} {\bibfnamefont
  {P.}~\bibnamefont {Goswami}}, \ and\ \bibinfo {author} {\bibfnamefont
  {S.}~\bibnamefont {Nakatsuji}},\ }\bibfield  {title} {\enquote {\bibinfo
  {title} {Giant anomalous Nernst effect and quantum-critical scaling in a
  ferromagnetic semimetal},}\ }\href@noop {} {\bibfield  {journal} {\bibinfo
  {journal} {Nature Phys.}\ }\textbf {\bibinfo {volume} {14}},\ \bibinfo
  {pages} {1119} (\bibinfo {year} {2016})}\BibitemShut {NoStop}%
\bibitem [{\citenamefont {Huang}\ \emph {et~al.}(2012)\citenamefont {Huang},
  \citenamefont {Fan}, \citenamefont {Qu}, \citenamefont {Chen}, \citenamefont
  {Wang}, \citenamefont {Wu}, \citenamefont {Chen}, \citenamefont {Xiao},\ and\
  \citenamefont {Chien}}]{HuangPRL2012}%
  \BibitemOpen
  \bibfield  {author} {\bibinfo {author} {\bibfnamefont {S.~Y.}\ \bibnamefont
  {Huang}}, \bibinfo {author} {\bibfnamefont {X.}~\bibnamefont {Fan}}, \bibinfo
  {author} {\bibfnamefont {D.}~\bibnamefont {Qu}}, \bibinfo {author}
  {\bibfnamefont {Y.~P.}\ \bibnamefont {Chen}}, \bibinfo {author}
  {\bibfnamefont {W.~G.}\ \bibnamefont {Wang}}, \bibinfo {author}
  {\bibfnamefont {J.}~\bibnamefont {Wu}}, \bibinfo {author} {\bibfnamefont
  {T.~Y.}\ \bibnamefont {Chen}}, \bibinfo {author} {\bibfnamefont {J.~Q.}\
  \bibnamefont {Xiao}}, \ and\ \bibinfo {author} {\bibfnamefont {C.~L.}\
  \bibnamefont {Chien}},\ }\bibfield  {title} {\enquote {\bibinfo {title}
  {Transport magnetic proximity effects in platinum},}\ }\href@noop {}
  {\bibfield  {journal} {\bibinfo  {journal} {Phys. Rev. Lett.}\ }\textbf
  {\bibinfo {volume} {109}},\ \bibinfo {pages} {107204} (\bibinfo {year}
  {2012})}\BibitemShut {NoStop}%
\bibitem [{\citenamefont {Kikkawa}\ \emph {et~al.}(2013)\citenamefont
  {Kikkawa}, \citenamefont {Uchida}, \citenamefont {Daimon}, \citenamefont
  {Shiomi}, \citenamefont {Adachi}, \citenamefont {Qui}, \citenamefont {Hou},
  \citenamefont {Jin}, \citenamefont {Maekawa},\ and\ \citenamefont
  {Saitoh}}]{KikkawaPRB2013}%
  \BibitemOpen
  \bibfield  {author} {\bibinfo {author} {\bibfnamefont {T.}~\bibnamefont
  {Kikkawa}}, \bibinfo {author} {\bibfnamefont {K.}~\bibnamefont {Uchida}},
  \bibinfo {author} {\bibfnamefont {S.}~\bibnamefont {Daimon}}, \bibinfo
  {author} {\bibfnamefont {Y.}~\bibnamefont {Shiomi}}, \bibinfo {author}
  {\bibfnamefont {H.}~\bibnamefont {Adachi}}, \bibinfo {author} {\bibfnamefont
  {Z.}~\bibnamefont {Qui}}, \bibinfo {author} {\bibfnamefont {D.}~\bibnamefont
  {Hou}}, \bibinfo {author} {\bibfnamefont {X.-F.}\ \bibnamefont {Jin}},
  \bibinfo {author} {\bibfnamefont {S.}~\bibnamefont {Maekawa}}, \ and\
  \bibinfo {author} {\bibfnamefont {E.}~\bibnamefont {Saitoh}},\ }\bibfield
  {title} {\enquote {\bibinfo {title} {Separation of longitudinal spin Seebeck
  effect from anomalous Nernst effect: Determination of origin of transverse
  thermoelectric voltage in metal/insulator junctions},}\ }\href@noop {}
  {\bibfield  {journal} {\bibinfo  {journal} {Phys. Rev. B}\ }\textbf {\bibinfo
  {volume} {88}},\ \bibinfo {pages} {214403} (\bibinfo {year}
  {2013})}\BibitemShut {NoStop}%
\bibitem [{\citenamefont {Miao}\ \emph {et~al.}(2016)\citenamefont {Miao},
  \citenamefont {Huang}, \citenamefont {Qu},\ and\ \citenamefont
  {Chien}}]{MiaoAIPAdvances2016}%
  \BibitemOpen
  \bibfield  {author} {\bibinfo {author} {\bibfnamefont {B.~F.}\ \bibnamefont
  {Miao}}, \bibinfo {author} {\bibfnamefont {S.~Y.}\ \bibnamefont {Huang}},
  \bibinfo {author} {\bibfnamefont {D.}~\bibnamefont {Qu}}, \ and\ \bibinfo
  {author} {\bibfnamefont {C.~L.}\ \bibnamefont {Chien}},\ }\bibfield  {title}
  {\enquote {\bibinfo {title} {Absence of anomalous Nernst effect in spin
  Seebeck effect fo Pt/YIG},}\ }\href@noop {} {\bibfield  {journal} {\bibinfo
  {journal} {AIP Advances}\ }\textbf {\bibinfo {volume} {6}},\ \bibinfo {pages}
  {015018} (\bibinfo {year} {2016})}\BibitemShut {NoStop}%
\bibitem [{\citenamefont {Agrawal}\ \emph {et~al.}(2014)\citenamefont
  {Agrawal}, \citenamefont {Vasyuchka}, \citenamefont {Serga}, \citenamefont
  {Kirihara}, \citenamefont {Pirro}, \citenamefont {Langer}, \citenamefont
  {Jungfleisch}, \citenamefont {Chumak}, \citenamefont {Papioannou},\ and\
  \citenamefont {Hillebrands}}]{AgrawalPRB2014}%
  \BibitemOpen
  \bibfield  {author} {\bibinfo {author} {\bibfnamefont {M.}~\bibnamefont
  {Agrawal}}, \bibinfo {author} {\bibfnamefont {V.~I.}\ \bibnamefont
  {Vasyuchka}}, \bibinfo {author} {\bibfnamefont {A.~A.}\ \bibnamefont
  {Serga}}, \bibinfo {author} {\bibfnamefont {A.}~\bibnamefont {Kirihara}},
  \bibinfo {author} {\bibfnamefont {P.}~\bibnamefont {Pirro}}, \bibinfo
  {author} {\bibfnamefont {T.}~\bibnamefont {Langer}}, \bibinfo {author}
  {\bibfnamefont {M.~B.}\ \bibnamefont {Jungfleisch}}, \bibinfo {author}
  {\bibfnamefont {A.~V.}\ \bibnamefont {Chumak}}, \bibinfo {author}
  {\bibfnamefont {E.~Th.}\ \bibnamefont {Papioannou}}, \ and\ \bibinfo {author}
  {\bibfnamefont {B.}~\bibnamefont {Hillebrands}},\ }\bibfield  {title}
  {\enquote {\bibinfo {title} {Role of bulk-magnon transport in the temporal
  evolution of the longitudinal spin-Seebeck effect},}\ }\href@noop {}
  {\bibfield  {journal} {\bibinfo  {journal} {Phys. Rev. B}\ }\textbf {\bibinfo
  {volume} {89}},\ \bibinfo {pages} {224414} (\bibinfo {year}
  {2014})}\BibitemShut {NoStop}%
\bibitem [{\citenamefont {Wang}\ \emph {et~al.}(2014)\citenamefont {Wang},
  \citenamefont {Du}, \citenamefont {Pu}, \citenamefont {Adur}, \citenamefont
  {Hammel},\ and\ \citenamefont {Yang}}]{WangPRL2014}%
  \BibitemOpen
  \bibfield  {author} {\bibinfo {author} {\bibfnamefont {H.~L.}\ \bibnamefont
  {Wang}}, \bibinfo {author} {\bibfnamefont {C.~H.}\ \bibnamefont {Du}},
  \bibinfo {author} {\bibfnamefont {Y.}~\bibnamefont {Pu}}, \bibinfo {author}
  {\bibfnamefont {R.}~\bibnamefont {Adur}}, \bibinfo {author} {\bibfnamefont
  {P.~C.}\ \bibnamefont {Hammel}}, \ and\ \bibinfo {author} {\bibfnamefont
  {F.~Y.}\ \bibnamefont {Yang}},\ }\bibfield  {title} {\enquote {\bibinfo
  {title} {Scaling of spin Hall angle in 3d, 4d, and 5d metals
  {Y}$_3${F}e$_5${O}$_{12}$/metal spin pumping},}\ }\href@noop {} {\bibfield
  {journal} {\bibinfo  {journal} {Phys. Rev. Lett.}\ }\textbf {\bibinfo
  {volume} {112}},\ \bibinfo {pages} {197201} (\bibinfo {year}
  {2014})}\BibitemShut {NoStop}%
\bibitem [{\citenamefont {Tan}\ \emph {et~al.}(2018)\citenamefont {Tan},
  \citenamefont {Huang}, \citenamefont {Ng}, \citenamefont {Wang},
  \citenamefont {Hasan}, \citenamefont {Duffin}, \citenamefont {Kumar},
  \citenamefont {Nijhuis}, \citenamefont {Lee},\ and\ \citenamefont
  {Ang}}]{TanAdvMater2018}%
  \BibitemOpen
  \bibfield  {author} {\bibinfo {author} {\bibfnamefont {W.~C.}\ \bibnamefont
  {Tan}}, \bibinfo {author} {\bibfnamefont {L.}~\bibnamefont {Huang}}, \bibinfo
  {author} {\bibfnamefont {R.~J.}\ \bibnamefont {Ng}}, \bibinfo {author}
  {\bibfnamefont {L.}~\bibnamefont {Wang}}, \bibinfo {author} {\bibfnamefont
  {D.~M.~N.}\ \bibnamefont {Hasan}}, \bibinfo {author} {\bibfnamefont {T.~J.}\
  \bibnamefont {Duffin}}, \bibinfo {author} {\bibfnamefont {K.~S.}\
  \bibnamefont {Kumar}}, \bibinfo {author} {\bibfnamefont {C.~A.}\ \bibnamefont
  {Nijhuis}}, \bibinfo {author} {\bibfnamefont {C.}~\bibnamefont {Lee}}, \ and\
  \bibinfo {author} {\bibfnamefont {K.-W.}\ \bibnamefont {Ang}},\ }\bibfield
  {title} {\enquote {\bibinfo {title} {A black phosphorus carbide infrared
  phototransistor},}\ }\href@noop {} {\bibfield  {journal} {\bibinfo  {journal}
  {Adv. Mater.}\ }\textbf {\bibinfo {volume} {30}},\ \bibinfo {pages} {1705039}
  (\bibinfo {year} {2018})}\BibitemShut {NoStop}%
\bibitem [{\citenamefont {Long}\ \emph {et~al.}(2016)\citenamefont {Long},
  \citenamefont {Liu}, \citenamefont {Wang}, \citenamefont {Gao}, \citenamefont
  {Xia}, \citenamefont {Luo}, \citenamefont {Wang}, \citenamefont {Zeng},
  \citenamefont {Fu}, \citenamefont {Xu}, \citenamefont {Zhou}, \citenamefont
  {Lv}, \citenamefont {Yao}, \citenamefont {Lu}, \citenamefont {Chen},
  \citenamefont {Ni}, \citenamefont {You}, \citenamefont {Zhang}, \citenamefont
  {Qin}, \citenamefont {Shi}, \citenamefont {Hu}, \citenamefont {Xing},\ and\
  \citenamefont {Miao}}]{LongNanoLett2016}%
  \BibitemOpen
  \bibfield  {author} {\bibinfo {author} {\bibfnamefont {M.}~\bibnamefont
  {Long}}, \bibinfo {author} {\bibfnamefont {E.}~\bibnamefont {Liu}}, \bibinfo
  {author} {\bibfnamefont {P.}~\bibnamefont {Wang}}, \bibinfo {author}
  {\bibfnamefont {A.}~\bibnamefont {Gao}}, \bibinfo {author} {\bibfnamefont
  {H.}~\bibnamefont {Xia}}, \bibinfo {author} {\bibfnamefont {W.}~\bibnamefont
  {Luo}}, \bibinfo {author} {\bibfnamefont {B.}~\bibnamefont {Wang}}, \bibinfo
  {author} {\bibfnamefont {J.}~\bibnamefont {Zeng}}, \bibinfo {author}
  {\bibfnamefont {Y.}~\bibnamefont {Fu}}, \bibinfo {author} {\bibfnamefont
  {K.}~\bibnamefont {Xu}}, \bibinfo {author} {\bibfnamefont {W.}~\bibnamefont
  {Zhou}}, \bibinfo {author} {\bibfnamefont {Y.}~\bibnamefont {Lv}}, \bibinfo
  {author} {\bibfnamefont {S.}~\bibnamefont {Yao}}, \bibinfo {author}
  {\bibfnamefont {M.}~\bibnamefont {Lu}}, \bibinfo {author} {\bibfnamefont
  {Y.}~\bibnamefont {Chen}}, \bibinfo {author} {\bibfnamefont {Z.}~\bibnamefont
  {Ni}}, \bibinfo {author} {\bibfnamefont {Y.}~\bibnamefont {You}}, \bibinfo
  {author} {\bibfnamefont {X.}~\bibnamefont {Zhang}}, \bibinfo {author}
  {\bibfnamefont {S.}~\bibnamefont {Qin}}, \bibinfo {author} {\bibfnamefont
  {Y.}~\bibnamefont {Shi}}, \bibinfo {author} {\bibfnamefont {W.}~\bibnamefont
  {Hu}}, \bibinfo {author} {\bibfnamefont {D.}~\bibnamefont {Xing}}, \ and\
  \bibinfo {author} {\bibfnamefont {F.}~\bibnamefont {Miao}},\ }\bibfield
  {title} {\enquote {\bibinfo {title} {Broadband photovoltaic detectors based
  on an atomically thin heterostructure},}\ }\href@noop {} {\bibfield
  {journal} {\bibinfo  {journal} {Nano Lett.}\ }\textbf {\bibinfo {volume}
  {16}},\ \bibinfo {pages} {2254} (\bibinfo {year} {2016})}\BibitemShut
  {NoStop}%
\bibitem [{\citenamefont {Buscema}\ \emph {et~al.}(2014)\citenamefont
  {Buscema}, \citenamefont {Groenendijk}, \citenamefont {Blanter},
  \citenamefont {Steele}, \citenamefont {van~der Zant},\ and\ \citenamefont
  {Castellanos-Gomez}}]{BuscemaNanoLett2014}%
  \BibitemOpen
  \bibfield  {author} {\bibinfo {author} {\bibfnamefont {M.}~\bibnamefont
  {Buscema}}, \bibinfo {author} {\bibfnamefont {D.~J.}\ \bibnamefont
  {Groenendijk}}, \bibinfo {author} {\bibfnamefont {S.~I.}\ \bibnamefont
  {Blanter}}, \bibinfo {author} {\bibfnamefont {G.~A.}\ \bibnamefont {Steele}},
  \bibinfo {author} {\bibfnamefont {H.~S.~J.}\ \bibnamefont {van~der Zant}}, \
  and\ \bibinfo {author} {\bibfnamefont {A.}~\bibnamefont
  {Castellanos-Gomez}},\ }\bibfield  {title} {\enquote {\bibinfo {title} {Fast
  and broadband photoresponse of few-layer black phosphorous field-effect
  transistors},}\ }\href@noop {} {\bibfield  {journal} {\bibinfo  {journal}
  {Nano Lett.}\ }\textbf {\bibinfo {volume} {14}},\ \bibinfo {pages} {3347}
  (\bibinfo {year} {2014})}\BibitemShut {NoStop}%
\bibitem [{\citenamefont {Wang}\ \emph {et~al.}(2015)\citenamefont {Wang},
  \citenamefont {Wang}, \citenamefont {Wang}, \citenamefont {Hu}, \citenamefont
  {Zhou}, \citenamefont {Guo}, \citenamefont {Huang}, \citenamefont {Sun},
  \citenamefont {Shen}, \citenamefont {Lin}, \citenamefont {Tang},
  \citenamefont {Liao}, \citenamefont {Jiang}, \citenamefont {Sun},
  \citenamefont {Meng}, \citenamefont {Chen}, \citenamefont {Lu},\ and\
  \citenamefont {Chu}}]{WangAdvMater2015}%
  \BibitemOpen
  \bibfield  {author} {\bibinfo {author} {\bibfnamefont {X.}~\bibnamefont
  {Wang}}, \bibinfo {author} {\bibfnamefont {P.}~\bibnamefont {Wang}}, \bibinfo
  {author} {\bibfnamefont {J.}~\bibnamefont {Wang}}, \bibinfo {author}
  {\bibfnamefont {W.}~\bibnamefont {Hu}}, \bibinfo {author} {\bibfnamefont
  {X.}~\bibnamefont {Zhou}}, \bibinfo {author} {\bibfnamefont {N.}~\bibnamefont
  {Guo}}, \bibinfo {author} {\bibfnamefont {H.}~\bibnamefont {Huang}}, \bibinfo
  {author} {\bibfnamefont {S.}~\bibnamefont {Sun}}, \bibinfo {author}
  {\bibfnamefont {H.}~\bibnamefont {Shen}}, \bibinfo {author} {\bibfnamefont
  {T.}~\bibnamefont {Lin}}, \bibinfo {author} {\bibfnamefont {M.}~\bibnamefont
  {Tang}}, \bibinfo {author} {\bibfnamefont {L.}~\bibnamefont {Liao}}, \bibinfo
  {author} {\bibfnamefont {A.}~\bibnamefont {Jiang}}, \bibinfo {author}
  {\bibfnamefont {J.}~\bibnamefont {Sun}}, \bibinfo {author} {\bibfnamefont
  {X.}~\bibnamefont {Meng}}, \bibinfo {author} {\bibfnamefont {X.}~\bibnamefont
  {Chen}}, \bibinfo {author} {\bibfnamefont {W.}~\bibnamefont {Lu}}, \ and\
  \bibinfo {author} {\bibfnamefont {J.}~\bibnamefont {Chu}},\ }\bibfield
  {title} {\enquote {\bibinfo {title} {Ultrasensitive and broadband MoS$_2$
  photodetector driven by ferroelectrics},}\ }\href@noop {} {\bibfield
  {journal} {\bibinfo  {journal} {Adv. Mater.}\ }\textbf {\bibinfo {volume}
  {27}},\ \bibinfo {pages} {6575} (\bibinfo {year} {2015})}\BibitemShut
  {NoStop}%
\bibitem [{\citenamefont {Liu}\ \emph {et~al.}(2014)\citenamefont {Liu},
  \citenamefont {Chang}, \citenamefont {Norris},\ and\ \citenamefont
  {Zhong}}]{LiuNatureNano2014}%
  \BibitemOpen
  \bibfield  {author} {\bibinfo {author} {\bibfnamefont {C-H.}\ \bibnamefont
  {Liu}}, \bibinfo {author} {\bibfnamefont {Y-C.}\ \bibnamefont {Chang}},
  \bibinfo {author} {\bibfnamefont {T.~B.}\ \bibnamefont {Norris}}, \ and\
  \bibinfo {author} {\bibfnamefont {Z.}~\bibnamefont {Zhong}},\ }\bibfield
  {title} {\enquote {\bibinfo {title} {Graphene photodetectors with
  ultra-broadband and high responsivity at room temperature},}\ }\href@noop {}
  {\bibfield  {journal} {\bibinfo  {journal} {Nature Nanotech.}\ }\textbf
  {\bibinfo {volume} {9}},\ \bibinfo {pages} {273} (\bibinfo {year}
  {2014})}\BibitemShut {NoStop}%
\bibitem [{\citenamefont {Liu}\ \emph {et~al.}(2015)\citenamefont {Liu},
  \citenamefont {Wang}, \citenamefont {Wang}, \citenamefont {Wang},
  \citenamefont {Flahaut}, \citenamefont {Liu}, \citenamefont {Li},
  \citenamefont {Wang}, \citenamefont {Xu}, \citenamefont {Shi},\ and\
  \citenamefont {Zhang}}]{LiuNatureComm2015}%
  \BibitemOpen
  \bibfield  {author} {\bibinfo {author} {\bibfnamefont {Y.}~\bibnamefont
  {Liu}}, \bibinfo {author} {\bibfnamefont {F.}~\bibnamefont {Wang}}, \bibinfo
  {author} {\bibfnamefont {X.}~\bibnamefont {Wang}}, \bibinfo {author}
  {\bibfnamefont {X.}~\bibnamefont {Wang}}, \bibinfo {author} {\bibfnamefont
  {E.}~\bibnamefont {Flahaut}}, \bibinfo {author} {\bibfnamefont
  {X.}~\bibnamefont {Liu}}, \bibinfo {author} {\bibfnamefont {Y.}~\bibnamefont
  {Li}}, \bibinfo {author} {\bibfnamefont {X.}~\bibnamefont {Wang}}, \bibinfo
  {author} {\bibfnamefont {Y.}~\bibnamefont {Xu}}, \bibinfo {author}
  {\bibfnamefont {Y.}~\bibnamefont {Shi}}, \ and\ \bibinfo {author}
  {\bibfnamefont {R.}~\bibnamefont {Zhang}},\ }\bibfield  {title} {\enquote
  {\bibinfo {title} {Planar carbon nanotube-graphene hybrid films for
  high-performance broadband photodetectors},}\ }\href@noop {} {\bibfield
  {journal} {\bibinfo  {journal} {Nature Comm.}\ }\textbf {\bibinfo {volume}
  {6}},\ \bibinfo {pages} {8589} (\bibinfo {year} {2015})}\BibitemShut
  {NoStop}%
\bibitem [{\citenamefont {Bandurin}\ \emph {et~al.}(2018)\citenamefont
  {Bandurin}, \citenamefont {Svintsov}, \citenamefont {Gayduchenko},
  \citenamefont {Xu}, \citenamefont {Principi}, \citenamefont {Moskotin},
  \citenamefont {Tretyakov}, \citenamefont {Yagodkin}, \citenamefont {Zhukov},
  \citenamefont {Taniguchi}, \citenamefont {Watanabe}, \citenamefont
  {Grigorieva}, \citenamefont {Polini}, \citenamefont {Goltsman}, \citenamefont
  {Geim},\ and\ \citenamefont {Fedorov}}]{BandurinNatureComm2018}%
  \BibitemOpen
  \bibfield  {author} {\bibinfo {author} {\bibfnamefont {D.~A.}\ \bibnamefont
  {Bandurin}}, \bibinfo {author} {\bibfnamefont {D.}~\bibnamefont {Svintsov}},
  \bibinfo {author} {\bibfnamefont {I.}~\bibnamefont {Gayduchenko}}, \bibinfo
  {author} {\bibfnamefont {S.~G.}\ \bibnamefont {Xu}}, \bibinfo {author}
  {\bibfnamefont {A.}~\bibnamefont {Principi}}, \bibinfo {author}
  {\bibfnamefont {M.}~\bibnamefont {Moskotin}}, \bibinfo {author}
  {\bibfnamefont {I.}~\bibnamefont {Tretyakov}}, \bibinfo {author}
  {\bibfnamefont {D.}~\bibnamefont {Yagodkin}}, \bibinfo {author}
  {\bibfnamefont {S.}~\bibnamefont {Zhukov}}, \bibinfo {author} {\bibfnamefont
  {T.}~\bibnamefont {Taniguchi}}, \bibinfo {author} {\bibfnamefont
  {K.}~\bibnamefont {Watanabe}}, \bibinfo {author} {\bibfnamefont {I.~V.}\
  \bibnamefont {Grigorieva}}, \bibinfo {author} {\bibfnamefont
  {M.}~\bibnamefont {Polini}}, \bibinfo {author} {\bibfnamefont {G.~N.}\
  \bibnamefont {Goltsman}}, \bibinfo {author} {\bibfnamefont {A.~K.}\
  \bibnamefont {Geim}}, \ and\ \bibinfo {author} {\bibfnamefont
  {G.}~\bibnamefont {Fedorov}},\ }\bibfield  {title} {\enquote {\bibinfo
  {title} {Resonant terahertz detection using graphene plasmons},}\ }\href@noop
  {} {\bibfield  {journal} {\bibinfo  {journal} {Nature Comm.}\ }\textbf
  {\bibinfo {volume} {9}},\ \bibinfo {pages} {5392} (\bibinfo {year}
  {2018})}\BibitemShut {NoStop}%
\bibitem [{\citenamefont {Yuasa}\ \emph {et~al.}(2018)\citenamefont {Yuasa},
  \citenamefont {Nakata}, \citenamefont {Nakamura},\ and\ \citenamefont
  {Kurokawa}}]{YuasaJPhysD2018}%
  \BibitemOpen
  \bibfield  {author} {\bibinfo {author} {\bibfnamefont {H.}~\bibnamefont
  {Yuasa}}, \bibinfo {author} {\bibfnamefont {F.}~\bibnamefont {Nakata}},
  \bibinfo {author} {\bibfnamefont {R.}~\bibnamefont {Nakamura}}, \ and\
  \bibinfo {author} {\bibfnamefont {Y.}~\bibnamefont {Kurokawa}},\ }\bibfield
  {title} {\enquote {\bibinfo {title} {Spin Seebeck coefficient enhancement by
  using Ta$_{50}$W$_{50}$ alloy and YIG/Ru interface},}\ }\href@noop {}
  {\bibfield  {journal} {\bibinfo  {journal} {J. Phys. D: Appl. Phys.}\
  }\textbf {\bibinfo {volume} {51}},\ \bibinfo {pages} {134002} (\bibinfo
  {year} {2018})}\BibitemShut {NoStop}%
\bibitem [{\citenamefont {Nakata}\ \emph {et~al.}(2019)\citenamefont {Nakata},
  \citenamefont {Niimura}, \citenamefont {Kurokawa},\ and\ \citenamefont
  {Yuasa}}]{NakataJJAP2019}%
  \BibitemOpen
  \bibfield  {author} {\bibinfo {author} {\bibfnamefont {F.}~\bibnamefont
  {Nakata}}, \bibinfo {author} {\bibfnamefont {T.}~\bibnamefont {Niimura}},
  \bibinfo {author} {\bibfnamefont {Y.}~\bibnamefont {Kurokawa}}, \ and\
  \bibinfo {author} {\bibfnamefont {H.}~\bibnamefont {Yuasa}},\ }\bibfield
  {title} {\enquote {\bibinfo {title} {Spin Seebeck voltage enhancement by Mn
  system metals insertion at the interface between YIG and nonmagnetic
  layer},}\ }\href@noop {} {\bibfield  {journal} {\bibinfo  {journal} {Japanese
  Journal of Appl. Phys.}\ }\textbf {\bibinfo {volume} {58}},\ \bibinfo {pages}
  {SBBI04} (\bibinfo {year} {2019})}\BibitemShut {NoStop}%
\bibitem [{\citenamefont {Weiler}\ \emph {et~al.}(2013)\citenamefont {Weiler},
  \citenamefont {Althammer}, \citenamefont {Schreier}, \citenamefont {Lotze},
  \citenamefont {Pernpeintner}, \citenamefont {Meyer}, \citenamefont {Huebl},
  \citenamefont {Gross}, \citenamefont {Kamra}, \citenamefont {Xiao},
  \citenamefont {Chen}, \citenamefont {Jiao}, \citenamefont {Bauer},\ and\
  \citenamefont {Goennenwein}}]{WeilerPRL2013}%
  \BibitemOpen
  \bibfield  {author} {\bibinfo {author} {\bibfnamefont {M.}~\bibnamefont
  {Weiler}}, \bibinfo {author} {\bibfnamefont {M.}~\bibnamefont {Althammer}},
  \bibinfo {author} {\bibfnamefont {M.}~\bibnamefont {Schreier}}, \bibinfo
  {author} {\bibfnamefont {J.}~\bibnamefont {Lotze}}, \bibinfo {author}
  {\bibfnamefont {M.}~\bibnamefont {Pernpeintner}}, \bibinfo {author}
  {\bibfnamefont {S.}~\bibnamefont {Meyer}}, \bibinfo {author} {\bibfnamefont
  {H.}~\bibnamefont {Huebl}}, \bibinfo {author} {\bibfnamefont
  {R.}~\bibnamefont {Gross}}, \bibinfo {author} {\bibfnamefont
  {A.}~\bibnamefont {Kamra}}, \bibinfo {author} {\bibfnamefont
  {J.}~\bibnamefont {Xiao}}, \bibinfo {author} {\bibfnamefont {Y.-T.}\
  \bibnamefont {Chen}}, \bibinfo {author} {\bibfnamefont {H.}~\bibnamefont
  {Jiao}}, \bibinfo {author} {\bibfnamefont {G.~E.~W.}\ \bibnamefont {Bauer}},
  \ and\ \bibinfo {author} {\bibfnamefont {S.~T.~B.}\ \bibnamefont
  {Goennenwein}},\ }\bibfield  {title} {\enquote {\bibinfo {title}
  {Experimental test of the spin mixing interface conductivity concept},}\
  }\href@noop {} {\bibfield  {journal} {\bibinfo  {journal} {Phys. Rev. Lett.}\
  }\textbf {\bibinfo {volume} {111}},\ \bibinfo {pages} {176601} (\bibinfo
  {year} {2013})}\BibitemShut {NoStop}%
\bibitem [{\citenamefont {Aqeel}\ \emph {et~al.}(2014)\citenamefont {Aqeel},
  \citenamefont {Vera-Marun}, \citenamefont {van Wees},\ and\ \citenamefont
  {Palstra}}]{AqeelJAP2014}%
  \BibitemOpen
  \bibfield  {author} {\bibinfo {author} {\bibfnamefont {A.}~\bibnamefont
  {Aqeel}}, \bibinfo {author} {\bibfnamefont {I.~J.}\ \bibnamefont
  {Vera-Marun}}, \bibinfo {author} {\bibfnamefont {B.~J.}\ \bibnamefont {van
  Wees}}, \ and\ \bibinfo {author} {\bibfnamefont {T.~T.~M.}\ \bibnamefont
  {Palstra}},\ }\bibfield  {title} {\enquote {\bibinfo {title} {Surface
  sensitivity of the spin Seebeck effect},}\ }\href@noop {} {\bibfield
  {journal} {\bibinfo  {journal} {J. Appl. Phys.}\ }\textbf {\bibinfo {volume}
  {116}},\ \bibinfo {pages} {153705} (\bibinfo {year} {2014})}\BibitemShut
  {NoStop}%
\bibitem [{\citenamefont {Surabhi}\ \emph {et~al.}(2018)\citenamefont
  {Surabhi}, \citenamefont {Kim}, \citenamefont {Van}, \citenamefont {Quoc},
  \citenamefont {Kim}, \citenamefont {Lee}, \citenamefont {Kuchi},
  \citenamefont {Lee}, \citenamefont {Yoon}, \citenamefont {Choi},
  \citenamefont {Park},\ and\ \citenamefont
  {Jeong}}]{SurabhiScientificRep2018}%
  \BibitemOpen
  \bibfield  {author} {\bibinfo {author} {\bibfnamefont {S.}~\bibnamefont
  {Surabhi}}, \bibinfo {author} {\bibfnamefont {D.-J.}\ \bibnamefont {Kim}},
  \bibinfo {author} {\bibfnamefont {P.~C.}\ \bibnamefont {Van}}, \bibinfo
  {author} {\bibfnamefont {V.~D.}\ \bibnamefont {Quoc}}, \bibinfo {author}
  {\bibfnamefont {J.-K.}\ \bibnamefont {Kim}}, \bibinfo {author} {\bibfnamefont
  {S.~W.}\ \bibnamefont {Lee}}, \bibinfo {author} {\bibfnamefont
  {R.}~\bibnamefont {Kuchi}}, \bibinfo {author} {\bibfnamefont {J.-W.}\
  \bibnamefont {Lee}}, \bibinfo {author} {\bibfnamefont {S.-G.}\ \bibnamefont
  {Yoon}}, \bibinfo {author} {\bibfnamefont {J.}~\bibnamefont {Choi}}, \bibinfo
  {author} {\bibfnamefont {B.-G.}\ \bibnamefont {Park}}, \ and\ \bibinfo
  {author} {\bibfnamefont {J.-R.}\ \bibnamefont {Jeong}},\ }\bibfield  {title}
  {\enquote {\bibinfo {title} {Precise determination of the temperature
  gradients in laser-irradiated ultrathin magnetic layers for the analysis of
  thermal spin current},}\ }\href@noop {} {\bibfield  {journal} {\bibinfo
  {journal} {Sci.~Rep.}\ }\textbf {\bibinfo {volume} {8}},\ \bibinfo {pages}
  {11337} (\bibinfo {year} {2018})}\BibitemShut {NoStop}%
\bibitem [{\citenamefont {Scott}\ and\ \citenamefont
  {Page}(1977)}]{Scottpssb1977}%
  \BibitemOpen
  \bibfield  {author} {\bibinfo {author} {\bibfnamefont {G.~B.}\ \bibnamefont
  {Scott}}\ and\ \bibinfo {author} {\bibfnamefont {J.~L.}\ \bibnamefont
  {Page}},\ }\bibfield  {title} {\enquote {\bibinfo {title} {The absorption
  spectra of Y$_3$Fe$_5$O$_{12}$ and Y$_3$Ga$_5$O$_{12}$:Fe$^{3+}$ to 5.5
  eV},}\ }\href@noop {} {\bibfield  {journal} {\bibinfo  {journal} {phys. stat.
  sol. (b)}\ }\textbf {\bibinfo {volume} {79}},\ \bibinfo {pages} {203}
  (\bibinfo {year} {1977})}\BibitemShut {NoStop}%
\bibitem [{\citenamefont {Ellabban}\ \emph {et~al.}(2006)\citenamefont
  {Ellabban}, \citenamefont {Fally}, \citenamefont {Rupp},\ and\ \citenamefont
  {Kov\'{a}cs}}]{EllabbanOpticsExpr2006}%
  \BibitemOpen
  \bibfield  {author} {\bibinfo {author} {\bibfnamefont {M.~A.}\ \bibnamefont
  {Ellabban}}, \bibinfo {author} {\bibfnamefont {M.}~\bibnamefont {Fally}},
  \bibinfo {author} {\bibfnamefont {R.~A.}\ \bibnamefont {Rupp}}, \ and\
  \bibinfo {author} {\bibfnamefont {L.}~\bibnamefont {Kov\'{a}cs}},\ }\bibfield
   {title} {\enquote {\bibinfo {title} {Light-induced phase and amplitude
  gratings in centrosymmetric gadolinium gallium garnet doped with calcium},}\
  }\href@noop {} {\bibfield  {journal} {\bibinfo  {journal} {Optics Express}\
  }\textbf {\bibinfo {volume} {14}},\ \bibinfo {pages} {593} (\bibinfo {year}
  {2006})}\BibitemShut {NoStop}%
\bibitem [{\citenamefont {Zhang}\ \emph {et~al.}(1997)\citenamefont {Zhang},
  \citenamefont {Nagao}, \citenamefont {Kuwano},\ and\ \citenamefont
  {Ito}}]{ZhangJJAP1997}%
  \BibitemOpen
  \bibfield  {author} {\bibinfo {author} {\bibfnamefont {J.}~\bibnamefont
  {Zhang}}, \bibinfo {author} {\bibfnamefont {Y.}~\bibnamefont {Nagao}},
  \bibinfo {author} {\bibfnamefont {S.}~\bibnamefont {Kuwano}}, \ and\ \bibinfo
  {author} {\bibfnamefont {Y.}~\bibnamefont {Ito}},\ }\bibfield  {title}
  {\enquote {\bibinfo {title} {Microstructure and temperature coefficient of
  resistance of platinum films},}\ }\href@noop {} {\bibfield  {journal}
  {\bibinfo  {journal} {Japanese Journal of Appl. Phys.}\ }\textbf {\bibinfo
  {volume} {36}},\ \bibinfo {pages} {834} (\bibinfo {year} {1997})}\BibitemShut
  {NoStop}%
\bibitem [{\citenamefont {Gale}\ and\ \citenamefont
  {Totemeier}(2004)}]{Gale2004}%
  \BibitemOpen
  \bibfield  {author} {\bibinfo {author} {\bibfnamefont {W.~F.}\ \bibnamefont
  {Gale}}\ and\ \bibinfo {author} {\bibfnamefont {T.~C.}\ \bibnamefont
  {Totemeier}},\ }\bibfield  {title} {\enquote {\bibinfo {title} {General
  physical properties},}\ }in\ \href@noop {} {\emph {\bibinfo {booktitle}
  {Smithells Metals Reference Book}}}\ (\bibinfo  {publisher} {Elsevier},\
  \bibinfo {year} {2004})\ pp.\ \bibinfo {pages} {14--1--14--45}\BibitemShut
  {NoStop}%
\bibitem [{\citenamefont {Namba}(1970)}]{NambaJJAP1970}%
  \BibitemOpen
  \bibfield  {author} {\bibinfo {author} {\bibfnamefont {Y.}~\bibnamefont
  {Namba}},\ }\bibfield  {title} {\enquote {\bibinfo {title} {Resistivity and
  temperature coefficient of thin metal films with rough surface},}\
  }\href@noop {} {\bibfield  {journal} {\bibinfo  {journal} {Japanese Journal
  of Appl. Phys.}\ }\textbf {\bibinfo {volume} {9}},\ \bibinfo {pages} {1326}
  (\bibinfo {year} {1970})}\BibitemShut {NoStop}%
\bibitem [{\citenamefont {Sola}\ \emph {et~al.}(2015)\citenamefont {Sola},
  \citenamefont {Kuepferling}, \citenamefont {Basso}, \citenamefont {Pasquale},
  \citenamefont {Kikkawa}, \citenamefont {Uchida},\ and\ \citenamefont
  {Saitoh}}]{SolaJAP2015}%
  \BibitemOpen
  \bibfield  {author} {\bibinfo {author} {\bibfnamefont {A.}~\bibnamefont
  {Sola}}, \bibinfo {author} {\bibfnamefont {M.}~\bibnamefont {Kuepferling}},
  \bibinfo {author} {\bibfnamefont {V.}~\bibnamefont {Basso}}, \bibinfo
  {author} {\bibfnamefont {M.}~\bibnamefont {Pasquale}}, \bibinfo {author}
  {\bibfnamefont {T.}~\bibnamefont {Kikkawa}}, \bibinfo {author} {\bibfnamefont
  {K.}~\bibnamefont {Uchida}}, \ and\ \bibinfo {author} {\bibfnamefont
  {E.}~\bibnamefont {Saitoh}},\ }\bibfield  {title} {\enquote {\bibinfo {title}
  {Evaluation of thermal gradients in longitudinal spin Seebeck effect
  measurements},}\ }\href@noop {} {\bibfield  {journal} {\bibinfo  {journal}
  {J. Appl. Phys.}\ }\textbf {\bibinfo {volume} {117}},\ \bibinfo {pages}
  {17C510} (\bibinfo {year} {2015})}\BibitemShut {NoStop}%
\bibitem [{\citenamefont {Sola}\ \emph {et~al.}(2017)\citenamefont {Sola},
  \citenamefont {Bougiatioti}, \citenamefont {Kuepferling}, \citenamefont
  {Meier}, \citenamefont {Reiss}, \citenamefont {Pasquale}, \citenamefont
  {Kuschel},\ and\ \citenamefont {Basso}}]{SolaScientificReports2017}%
  \BibitemOpen
  \bibfield  {author} {\bibinfo {author} {\bibfnamefont {A.}~\bibnamefont
  {Sola}}, \bibinfo {author} {\bibfnamefont {P.}~\bibnamefont {Bougiatioti}},
  \bibinfo {author} {\bibfnamefont {M.}~\bibnamefont {Kuepferling}}, \bibinfo
  {author} {\bibfnamefont {D.}~\bibnamefont {Meier}}, \bibinfo {author}
  {\bibfnamefont {G.}~\bibnamefont {Reiss}}, \bibinfo {author} {\bibfnamefont
  {M.}~\bibnamefont {Pasquale}}, \bibinfo {author} {\bibfnamefont
  {T.}~\bibnamefont {Kuschel}}, \ and\ \bibinfo {author} {\bibfnamefont
  {V.}~\bibnamefont {Basso}},\ }\bibfield  {title} {\enquote {\bibinfo {title}
  {Longitudinal spin Seebeck coefficient: heat flux vs. temperature difference
  method},}\ }\href@noop {} {\bibfield  {journal} {\bibinfo  {journal}
  {Sci.~Rep.}\ }\textbf {\bibinfo {volume} {7}},\ \bibinfo {pages} {46752}
  (\bibinfo {year} {2017})}\BibitemShut {NoStop}%
\bibitem [{\citenamefont {Boukai}\ \emph {et~al.}(2008)\citenamefont {Boukai},
  \citenamefont {Bunimovich}, \citenamefont {Tahir-Kheli}, \citenamefont {Yu},
  \citenamefont {{Goddard III}},\ and\ \citenamefont
  {Heath}}]{BoukaiNature2008}%
  \BibitemOpen
  \bibfield  {author} {\bibinfo {author} {\bibfnamefont {A.~I.}\ \bibnamefont
  {Boukai}}, \bibinfo {author} {\bibfnamefont {Y.}~\bibnamefont {Bunimovich}},
  \bibinfo {author} {\bibfnamefont {J.}~\bibnamefont {Tahir-Kheli}}, \bibinfo
  {author} {\bibfnamefont {J.-K.}\ \bibnamefont {Yu}}, \bibinfo {author}
  {\bibfnamefont {W.~A.}\ \bibnamefont {{Goddard III}}}, \ and\ \bibinfo
  {author} {\bibfnamefont {J.~R.}\ \bibnamefont {Heath}},\ }\bibfield  {title}
  {\enquote {\bibinfo {title} {Silicon nanowires as efficient thermoelectric
  materials},}\ }\href@noop {} {\bibfield  {journal} {\bibinfo  {journal}
  {Nature}\ }\textbf {\bibinfo {volume} {451}},\ \bibinfo {pages} {168}
  (\bibinfo {year} {2008})}\BibitemShut {NoStop}%
\bibitem [{\citenamefont {Ye}\ \emph {et~al.}(2017)\citenamefont {Ye},
  \citenamefont {Wang}, \citenamefont {Li}, \citenamefont {Yan}, \citenamefont
  {Ramakrishna},\ and\ \citenamefont {Xu}}]{YeJMaterChemC2017}%
  \BibitemOpen
  \bibfield  {author} {\bibinfo {author} {\bibfnamefont {T.}~\bibnamefont
  {Ye}}, \bibinfo {author} {\bibfnamefont {X.}~\bibnamefont {Wang}}, \bibinfo
  {author} {\bibfnamefont {X.}~\bibnamefont {Li}}, \bibinfo {author}
  {\bibfnamefont {A.~Q.}\ \bibnamefont {Yan}}, \bibinfo {author} {\bibfnamefont
  {S.}~\bibnamefont {Ramakrishna}}, \ and\ \bibinfo {author} {\bibfnamefont
  {J.}~\bibnamefont {Xu}},\ }\bibfield  {title} {\enquote {\bibinfo {title}
  {Ultra-high Seebeck coefficient and low thermal conductivity of a
  centimeter-sized perovskite single crystal acquired by a modified fast growth
  method},}\ }\href@noop {} {\bibfield  {journal} {\bibinfo  {journal} {J.
  Mater. Chem. C}\ }\textbf {\bibinfo {volume} {5}},\ \bibinfo {pages} {1255}
  (\bibinfo {year} {2017})}\BibitemShut {NoStop}%
\bibitem [{\citenamefont {Seki}\ \emph {et~al.}(2008)\citenamefont {Seki},
  \citenamefont {Hasegawa}, \citenamefont {Mitani}, \citenamefont {Takahashi},
  \citenamefont {Imamura}, \citenamefont {Maekawa}, \citenamefont {Nitta},\
  and\ \citenamefont {Takanashi}}]{SekiNatureMater2008}%
  \BibitemOpen
  \bibfield  {author} {\bibinfo {author} {\bibfnamefont {T.}~\bibnamefont
  {Seki}}, \bibinfo {author} {\bibfnamefont {Y.}~\bibnamefont {Hasegawa}},
  \bibinfo {author} {\bibfnamefont {S.}~\bibnamefont {Mitani}}, \bibinfo
  {author} {\bibfnamefont {S.}~\bibnamefont {Takahashi}}, \bibinfo {author}
  {\bibfnamefont {H.}~\bibnamefont {Imamura}}, \bibinfo {author} {\bibfnamefont
  {S.}~\bibnamefont {Maekawa}}, \bibinfo {author} {\bibfnamefont
  {J.}~\bibnamefont {Nitta}}, \ and\ \bibinfo {author} {\bibfnamefont
  {K.}~\bibnamefont {Takanashi}},\ }\bibfield  {title} {\enquote {\bibinfo
  {title} {Giant spin hall effect in perpendicularly spin-polarized FePt/Au
  devices},}\ }\href@noop {} {\bibfield  {journal} {\bibinfo  {journal} {Nature
  Mater.}\ }\textbf {\bibinfo {volume} {7}},\ \bibinfo {pages} {125} (\bibinfo
  {year} {2008})}\BibitemShut {NoStop}%
\end{thebibliography}

%

\end{document}